\documentclass[aps,prX,twocolumn,showpacs,amsmath,amssymb,superscriptaddress,flushbottom,noraggedfooter]{revtex4-1}

\usepackage{graphicx}
\usepackage{subfigure}
\usepackage{amssymb}
\usepackage{multirow}
\usepackage{upgreek}
\usepackage[breaklinks,ps2pdf,pdftitle={Precision measurement of Rb-87 scattering lengths (PRA)}, pdfauthor={Mikhail Egorov}]{hyperref}
\usepackage[usenames]{color}

\newcommand{\ket}[1]{\mbox{\ensuremath{| #1 \rangle}}}
\newcommand{\Rb}{$^{87}$Rb}
\newcommand{\figref}[1]{Fig.~\ref{#1}}
\newcommand{\tf}{\textrm{\tiny{TF}}}

\renewcommand\Im{\operatorname{Im}}

\begin{document}

\title{Measurement of $s$-wave scattering lengths in a two-component Bose-Einstein condensate}

\author{M.~Egorov}
\email{mikhail.egorov@monash.edu}
\altaffiliation{Now at School of Physics, Monash University, Victoria 3800, Australia}
\author{B.~Opanchuk}
\author{P.~Drummond}
\author{B.V.~Hall}
\author{P.~Hannaford}
\author{A.~I.~Sidorov}
\email{asidorov@swin.edu.au}
\affiliation{
ARC Centre of Excellence for Quantum-Atom Optics and Centre for Atom Optics and Ultrafast Spectroscopy,\\
Swinburne University of Technology, Melbourne 3122, Australia
}

\begin{abstract}
	We use collective oscillations of a two-component Bose-Einstein condensate (2CBEC) of \Rb~atoms prepared in the internal states $\ket{1}\equiv\ket{F=1,~m_F=-1}$ and $\ket{2}\equiv\ket{F=2,~m_F=1}$ for the precision measurement of the interspecies scattering length $a_{12}$ with a relative uncertainty of $1.6\times 10^{-4}$.
	We show that in a cigar-shaped trap the three-dimensional (3D) dynamics of a component with a small relative population can be conveniently described by a one-dimensional (1D) Schr\"{o}dinger equation for an effective harmonic oscillator.
	The frequency of the collective oscillations is defined by the axial trap frequency and the ratio $a_{12}/a_{11}$, where $a_{11}$ is the intraspecies scattering length of a highly populated component $1$, and is largely decoupled from the scattering length $a_{22}$, the total atom number and loss terms.
	By fitting numerical simulations of the coupled Gross-Pitaevskii equations to the recorded temporal evolution of the axial width we obtain the value $a_{12}=98.006(16)\,a_0$, where $a_0$ is the Bohr radius.
	Our reported value is in a reasonable agreement with the theoretical prediction $a_{12}=98.13(10)\,a_0$ but deviates significantly from the previously measured value $a_{12}=97.66\,a_0$~\cite{Mertes07} which is commonly used in the characterisation of spin dynamics in degenerate \Rb~atoms.
	Using Ramsey interferometry of the 2CBEC we measure the scattering length $a_{22}=95.44(7)\,a_0$ which also deviates from the previously reported value $a_{22}=95.0\,a_0$~\cite{Mertes07}.
	We characterise two-body losses for the component $2$ and obtain the loss coefficients ${\gamma_{12}=1.51(18)\times10^{-14}~\textrm{cm}^3/\textrm{s}}$ and ${\gamma_{22}=8.1(3)\times10^{-14}~\textrm{cm}^3/\textrm{s}}$.
\end{abstract}


\date{\today}

\maketitle

\section{Introduction}

Collisional interactions in dilute ultracold gases play an important role in the dynamics of Bose-Einstein condensates (BECs), formation of molecules and shifts of resonance frequencies.
Binary collisions of atoms can be divided into inelastic and elastic collisions.
Inelastic collisions lead to the change of a hyperfine state or spin flips of the colliding atoms.
At sufficiently low temperatures the elastic collisions of ultracold bosonic atoms can be adequately described by a single parameter, the $s$-wave scattering length $a$, through the corresponding interaction strength~\cite{Pitaevskii03}.

Precise knowledge of the values of the scattering length is required for reliable modeling of BEC dynamics and spin squeezing, the accurate evaluation of collisional shifts in atomic clocks, and for spin gradient thermometry at sub-nanokelvin temperatures~\cite{Weld10}.
It can also be used to verify theoretical models of inter-atomic potentials~\cite{Verhaar09}.
In general, it is difficult to carry out precision measurements of a scattering length in a single species ensemble.
Binary mixtures in the form of either two condensates of different atomic species or two-component Bose-Einstein condensates (2CBEC) provide more opportunities for the accurate measurement of these collisional properties.
Two-component BECs are defined as a mixture of two different spin- or hyperfine states of the same condensed species.
In either case collisions involve intraspecies ($a_{11}$ and $a_{22}$) and interspecies ($a_{12}$) $s$-wave scattering.

Due to the particular properties of the singlet and triplet inter-atomic potentials~\cite{vanKempen02} the $s$-wave scattering lengths of \Rb~atoms in the ground hyperfine states with $F=1$ and $F=2$ are very close to each other (the maximum difference is around $5\%$).
These two states in \Rb, $\ket{1}\equiv\ket{F=1,m_F=-1}$ and $\ket{2}\equiv\ket{F=2,m_F=1}$, are potentially useful for future applications of ultracold or condensed atoms because they are magnetically trappable and their differential first-order Zeeman shift is cancelled at a rather low magnetic field strength $3.228$~G~\cite{Harber02}.
As a result, coherent superpositions of these two states are largely insensitive to magnetic field noise.
Under appropriate conditions very long coherence times have been reported for trapped non-condensed atomic ensembles~\cite{Deutsch10} and a 2CBEC~\cite{Egorov11}.
This makes these states of great interest for on-chip atomic clocks and interferometric applications, since the collisional frequency shift in a trapped atomic clock with equal population of the two states is proportional to the difference $(a_{11}-a_{22})$ between the intraspecies scattering lengths.
The coherent superposition of these states and subsequent nonlinear evolution were recently used in a spin-squeezing experiment~\cite{Riedel10}.

In this paper we demonstrate a new method for precision measurement of the interspecies scattering length $a_{12}$ using collective oscillations in a 2CBEC.
It was previously proposed~\cite{Pu98} that the interspecies coupling has a dramatic effect on the collective excitation spectrum.
Our method is largely decoupled from the $a_{22}$ scattering length, the total atom number and loss terms.
In the course of our study we measured the two-body loss coefficients for the states $\ket{1}$ and $\ket{2}$ in \Rb~and obtained values that are significantly different from previous measurements~\cite{Mertes07,Tojo09}.
We also carried out a measurement of the intraspecies scattering length $a_{22}$ using Ramsey interferometry of a trapped 2CBEC, a known theoretical value of $a_{11}$ and our measured value of $a_{12}$.

In Sec.~\ref{sec:1d-effective-treatment}, we use the quantum least-action principle for a 2CBEC trapped in a cigar-shaped harmonic potential to derive a 1D Schr\"{o}dinger equation describing collective oscillations of the component $2$.
The oscillation frequency is defined by the axial trap frequency and the ratio of the scattering lengths $a_{12}/a_{11}$.
In Sec.~\ref{sec:gpe-simulations}, we use a full three-dimensional simulation of the coupled Gross-Pitaevskii equations (GPE) with loss terms to confirm that the frequency of the collective oscillations in the approximation of a small relative population of the component $2$ ($N_2\ll N_1$), largely depends on the value of $a_{12}$ (relative to $a_{11}$) and has a very weak dependence on the $a_{22}$ value and the total atom number $N = N_1 + N_2$ (for $N > 3\times 10^4$).
We describe our experimental setup and the characterization of the trap frequencies in Sec.~\ref{sec:experimental-setup}.
Section~\ref{sec:convergence} describes the converging analysis sequence which we use to obtain values of the $s$-wave scattering lengths $a_{12}$, $a_{22}$ and the two-body loss coefficients $\gamma_{12}$ and $\gamma_{22}$.
In Sec.~\ref{sec:loss-coefficients}, we present the results of our measurement of the two-body loss coefficients.
We describe in detail our measurements of the scattering length $a_{12}$ using collective oscillations in a two-component BEC and $a_{22}$ using Ramsey interferometry in Sections~\ref{sec:a12-measurements} and~\ref{sec:a22-measurement}, respectively.
We compare our results with previous experimental and theoretical investigations in Sec.~\ref{sec:discussion}.

\section{One-dimensional effective single-component treatment}
\label{sec:1d-effective-treatment}

We consider the dynamics of a two-component Bose-Einstein condensate initially prepared in the internal quantum state $\ket{1}$ (component $1$) and
trapped in a cigar-shaped, axially symmetric harmonic potential ${V=m\omega_z^2 z^2/2 + m\omega_r^2 r^2/2}$, where ${r^2=x^2+y^2}$.
Electromagnetic radiation can transfer a variable portion of the condensate to another internal quantum state $\ket{2}$ (component $2$) which is also trapped
in the same potential $V$.
The transfer modifies mean-field interactions and initiates a dynamical evolution of the two condensate wave functions $\Psi_1(\mathbf{r},t)$ and $\Psi_2(\mathbf{r},t)$ normalized to the atom numbers $N_1$ and $N_2$ in the components $1$ and $2$ respectively.
In this section, we obtain an analytic expression which substantiates decoupling of the $a_{12}$ measurement from the parameters $a_{22}$ and $N$.

We consider the case of a tight transverse confinement ($\omega_r\gg\omega_z$) where the 3D dynamics of the 2CBEC can be conveniently described by an effective 1D treatment.
We use the variational method and follow the procedure developed for single-component condensates~\cite{Salasnich02,Massignan03,Kamchatnov04}.
The action functional of a 2CBEC can be written as~\cite{Young-S10}
\begin{equation}
	S = \int \left(\mathcal{L}_1 + \mathcal{L}_2 - U_{12}\left|\Psi_1\right|^2\left|\Psi_2\right|^2\right)\,d^3\mathbf{r}\,dt,
	\label{eq:3d-action}
\end{equation}
where the Lagrangian density of the component $\ket{j}$ is
\begin{equation}
	\begin{split}
		\mathcal{L}_j & = i\frac{\hbar}2 \left( \Psi_j^*\frac{\partial}{\partial t}\Psi_j  - \Psi_j\frac{\partial}{\partial t}\Psi_j^* \right) \\
		& - \frac{\hbar^2}{2m} \left|\nabla\Psi_j\right|^2 - V\left|\Psi_j\right|^2 - \frac12 U_{jj} \left|\Psi_j\right|^4,
	\end{split}
	\label{eq:lagrangian-density}
\end{equation}
where $U_{ij}=4\pi\hbar^2 a_{ij}/m$ are the inter- and intra-component interaction strengths,
and $a_{ij}$ are the $s$-wave scattering lengths ($i,j = 1,2$).
Coupled three-dimensional GPE can be obtained as $\partial S/\partial\Psi_j^*=0$~\cite{Pitaevskii03}: 
\begin{equation}
	\begin{split}
		i\hbar\frac{\partial\Psi_1}{\partial t} &= \left[-\frac{\hbar^2\nabla^2}{2m} + V
			+ U_{11}\lvert\Psi_1\rvert^2 + U_{12}\lvert\Psi_2\rvert^2\right]\Psi_1,\\
		i\hbar\frac{\partial\Psi_2}{\partial t} &= \left[-\frac{\hbar^2\nabla^2}{2m} + V
			+ U_{12}\lvert\Psi_1\rvert^2 + U_{22}\lvert\Psi_2\rvert^2\right]\Psi_2.
	\end{split}
	\label{eq:cgpe}
\end{equation}

In order to reduce the 3D treatment to the 1D case, we factorize the wavefunctions in the form~\cite{Salasnich02}
\begin{equation}
	\Psi_j(\mathbf{r},t) = \phi_j \left(r, \sigma_j \left(z,t\right) \right) f_j \left(z,t\right),
	\label{eq:psi-decomposition}
\end{equation}
where $f_j$ is normalized to the atom number in component $j$ and $\phi_j$ is a Gaussian trial function normalized to unity
\begin{equation}
	\phi_j \left(r,\sigma_j\left(z,t\right)\right) =
		\frac1{\sqrt{\pi}\sigma_j\left(z,t\right)}
		e^{- \frac{r^2}{2\sigma_j \left(z,t\right)^2}}.
	\label{eq:radial-trial-wavefunction}
\end{equation}

The use of Gaussian trial functions for the radial dependence of the condensate density is justified in the limit of weak interactions when the BEC is one-dimensional~\cite{Perez-Garcia96,Salasnich02}.
A typical Thomas-Fermi~(TF) radius for the BEC in our experiments along the tight trap direction is $4~\mu\textrm{m}$,
four times larger than the size of the corresponding harmonic oscillator ground state.
However a Gaussian trial wavefunction is known to give consistent results for a 1D reduction even in the case of a TF radial profile of a BEC~\cite{Salasnich02}.
We assume that $\phi_{j}$ is slowly varying along the axial coordinate relative to the radial direction and
\begin{equation}
	\nabla^2 \phi_j \approx \left(
		\frac{\partial^2}{\partial x^2} + \frac{\partial^2}{\partial y^2}
	\right) \phi_j.
	\label{eq:radial-wavefunction-laplassian}
\end{equation}
Using the Euler-Lagrange equations ($\partial S/\partial f_j^* = 0$ and $\partial S/\partial\sigma_j = 0$) we obtain
\begin{equation}
	\begin{split}
		i\hbar \frac{\partial}{\partial t} f_1 = & \left[
			- \frac{\hbar^2}{2m} \frac{\partial^2}{\partial z^2} + \frac{m\omega_z^2 z^2}2 + \left(
				\frac{\hbar^2}{2m\sigma_1^2} + \frac{m\omega_r^2\sigma_1^2}2
			\right) \right.\\
		+ & \left.
			\frac{U_{11}}{2\pi\sigma_1^2} \left|f_1\right|^2 + \frac{U_{12}}{\pi\left(\sigma_1^2 + \sigma_2^2\right)}
			\left|f_2\right|^2
		\right] f_1,
	\end{split}
	\label{eq:lagrange-f1}
\end{equation}
\begin{equation}
	\begin{split}
		- \frac{\hbar^2}{2m} \sigma_1^{-3} + \frac{m\omega_r^2\sigma_1}2 & - \frac12 \frac{U_{11}}{2\pi\sigma_1^3} \left|f_1\right|^2 \\
		& - \frac{U_{12}\sigma_1}{\pi\left(\sigma_1^2+\sigma_2^2\right)^2} \left|f_{2}\right|^2 = 0;
	\end{split}
	\label{eq:lagrange-sigma1}
\end{equation}
\begin{equation}
	\begin{split}
		i\hbar \frac{\partial}{\partial t} f_2 = & \left[
			-\frac{\hbar^2}{2m} \frac{\partial^2}{\partial z^2} + \frac{m\omega_z^2 z^2}2 + \left(
				\frac{\hbar^2}{2m\sigma_2^2} + \frac{m\omega_r^2\sigma_2^2}2
			\right) \right.\\
		+ & \left.
			\frac{U_{22}}{2\pi\sigma_2^2} \left|f_{2}\right|^2 + \frac{U_{12}}{\pi\left(\sigma_1^2+\sigma_2^2\right)}
			\left|f_1\right|^2
		\right] f_2,
	\end{split}
	\label{eq:lagrange-f2}
\end{equation}
\begin{equation}
	\begin{split}
		-\frac{\hbar^2}{2m} \sigma_2^{-3} + \frac{m\omega_r^2\sigma_2}2 & - \frac12\frac{U_{22}}{2\pi\sigma_2^3} \left|f_2\right|^2 \\
		& - \frac{U_{12}\sigma_2}{\pi\left(\sigma_1^2 + \sigma_2^2\right)^2} \left|f_1\right|^2 = 0.
	\end{split}
	\label{eq:lagrange-sigma2}
\end{equation}

For the case of the transfer of a small atom number $N_2$ to state $\ket{2}$ ($\left|f_{2}\right|\ll\left|f_{1}\right|$) and a large atom number $N_1$ in state $\ket{1}$ ($\left|f_1\right|^2 \gg (2\,a_{11})^{-1}$) we find
\begin{eqnarray}
	\sigma_1^2 & = & \frac{\hbar}{m\omega_{r}} \sqrt{2 a_{11}}\left|f_{1}\right|,\label{eq:sigma1-high-n}\\
	\sigma_2^2 & = & \sigma_{1}^{2}\left(2\sqrt{\frac{a_{12}}{a_{11}}}-1\right). \label{eq:sigma2-vs-sigma1}
\end{eqnarray}
In the approximation $\left|f_2\right|^2/\left|f_1\right|^2\ll 1$, the density in component $\ket{1}$ does not change and is given in the TF approximation by
\begin{eqnarray}
	\left|f_1\right| & = & \frac{\sqrt2}{3\hbar\omega_r \sqrt{a_{11}}} \left(\mu - \frac{m\omega_z^2 z^2}2\right), \label{eq:tf-component-1} \\
	\mu & = & \left(\frac{135 N a_{11}\hbar^2\omega_r^2\omega_z\sqrt{m}}{2^{\frac{11}{2}}}\right)^{\frac{2}{5}}, \label{eq:effective-chemical-potential}
\end{eqnarray}
where $\mu$ is the effective 1D chemical potential of component $\ket{1}$.
Equation~\ref{eq:tf-component-1} is valid for $\lvert z\rvert < r_{\tf} = (2\mu / m\omega_z^2)^{1/2}$, otherwise $\lvert f_1\rvert=0$.
We substitute Eq.~\ref{eq:tf-component-1} instead of $\left|f_1\right|$ for simplicity.
However this implies a certain limitation on $r_{\tf}$ for which the
analytical solution is valid, which is discussed after the solution is obtained~(Eq.~\ref{eq:rtf-criterion}).
Now the effective 1D equation for component $\ket{2}$ is
\begin{equation}
	i\hbar\frac{\partial}{\partial t}f_2 = \left[- \frac{\hbar^2}{2m} \frac{\partial^2}{\partial z^2} + \frac{m\omega_{\textrm{eff}}^2 z^2}2
	+ \mu_{\textrm{eff}} \right] f_2,
	\label{eq:effective-ho-se}
\end{equation}
where
\begin{eqnarray}
	\omega_{\textrm{eff}} &=& \frac2{\sqrt{3}} \sqrt{1 - \sqrt{\frac{a_{12}}{a_{11}}}}\,\omega_z,\label{eq:effective-ho-w}\\
	\mu_{\textrm{eff}} &=& \frac{\mu}3 \left(4\sqrt{\frac{a_{12}}{a_{11}}} - 1 \right).\label{eq:effective-ho-mu}
\end{eqnarray}
Apart from the constant term $\mu_{\textrm{eff}}$, this is the Schr\"{o}dinger equation for a harmonic oscillator.
If the superposition of states $\ket{1}$ and $\ket{2}$ is prepared by a pulse with area $\theta$,
the 1D wavefunction of state $\ket{2}$ is expressed in terms of a TF profile for state $\ket{1}$ as
${f_2(z,0) = \sin^2(\theta/2)\,f_1(z,0)}$.
Therefore, the solution of Eq.~\ref{eq:effective-ho-se} takes the form
\begin{equation}
	\begin{split}
		f_2(z,t) & = e^{-i \mu_{\textrm{eff}} t / \hbar} \sum\limits_{k=0}^{\infty} \left[
			\vphantom{\int}
			e^{-i\omega_{\textrm{eff}}\left(2k + \frac12\right)\,t} \psi_{\textrm{ho}}(2k,z)\right.\\
		& \left.\times \int\psi_{\textrm{ho}}(2k,\xi)\,f_2(\xi,0)\,d\xi\right],
	\end{split}
	\label{eq:effective-ho-solution}
\end{equation}
where only even harmonic oscillator eigenstates $\psi_{\textrm{ho}}(2k, z)$ contribute to the solution because the wavefunction $f_2(z,0)$ is
symmetric about $z=0$.
Equation~\ref{eq:effective-ho-solution} is periodic in such a way that
\begin{equation}
	f_2(z,t + n/f_c) = e^{-i \mu_{\textrm{eff}} n / (\hbar f_c)} f_2(z,t),\quad n\in\mathbb{Z},
\end{equation}
where $f_c=2\times \omega_{\textrm{eff}}/2\pi$ gives the frequency of the collective oscillations
\begin{equation}
	f_{\textrm{c}}= \frac{4f_z}{\sqrt3} \sqrt{1 - \sqrt{\frac{a_{12}}{a_{11}}}}.
	\label{eq:collective-oscillations-frequency}
\end{equation}

The effective harmonic potential in equation~\ref{eq:effective-ho-se} acts only within the size of the BEC whose density is non-zero when
$\left|z\right|<r_{\tf}$.
Therefore, Eq.~\ref{eq:effective-ho-se} is valid only when the characteristic size of the relevant harmonic oscillator eigenstates is less than $r_{\tf}$.
As a criterion, we require $N$ to be large enough to make the first two even eigenstates of the harmonic oscillator smaller than $r_{\tf}$, so that
\begin{equation}
	r_{\tf}^2 \gg \frac{5\hbar}{m\omega_{\textrm{eff}}},\quad\textrm{or}
	\label{eq:rtf-criterion}
\end{equation}
\begin{equation}
	N \gg \frac{2.3}{a_{11} \omega_r^2} \sqrt{\frac{\hbar\omega_z^3}m} \left(1 - \sqrt{\frac{a_{12}}{a_{11}}}\right)^{-\frac54}. \label{eq:large-N-criterion}
\end{equation}
In our experiments the critical value of $N$ for this criterion is $5\times10^3$, and we choose $N$ to be at least $10$ times larger than this value.

Thus the transfer of a small atom number from state $\ket{1}$ to state $\ket{2}$ initiates collective oscillations of component $2$
along the axial coordinate $z$ with frequency $f_c$ which is independent of the total atom number $N$ (providing the condition of Eq.~\ref{eq:large-N-criterion} holds),
the intra-component scattering length $a_{22}$ and the transferred fraction.
The evolution of the 1D density of component $2$ ($n_2 = f_2^* f_2$) is presented in~\figref{fig:density-analytic} and clearly shows periodic compressions of component
$2$ with frequency $2.91$~Hz ($a_{12}/a_{11} = 98.006/100.40,~f_z = 11.507~\textrm{Hz}$).
Accurate measurements of the axial trap frequency and the frequency of the collective oscillations make precision measurements of
the ratio $a_{12}/a_{11}$ possible.
When $a_{12} < a_{11}$ component $2$ contracts periodically.
For $a_{12} > a_{11}$ the dynamics of component $2$ becomes unstable.

\section{GPE simulations of collective oscillations of component $2$}
\label{sec:gpe-simulations}

\begin{figure}
	\begin{center}
		\subfigure{\label{fig:density-analytic}\includegraphics{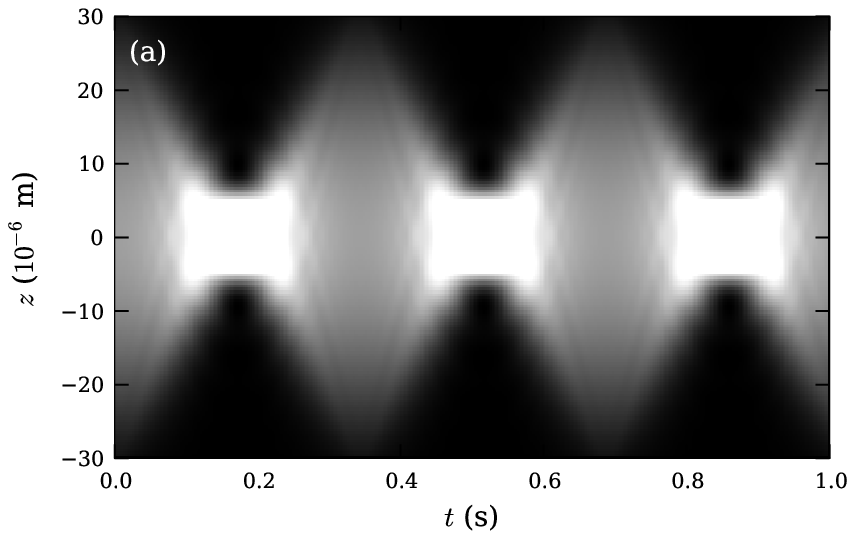}}
		\subfigure{\label{fig:density-gpe}\includegraphics{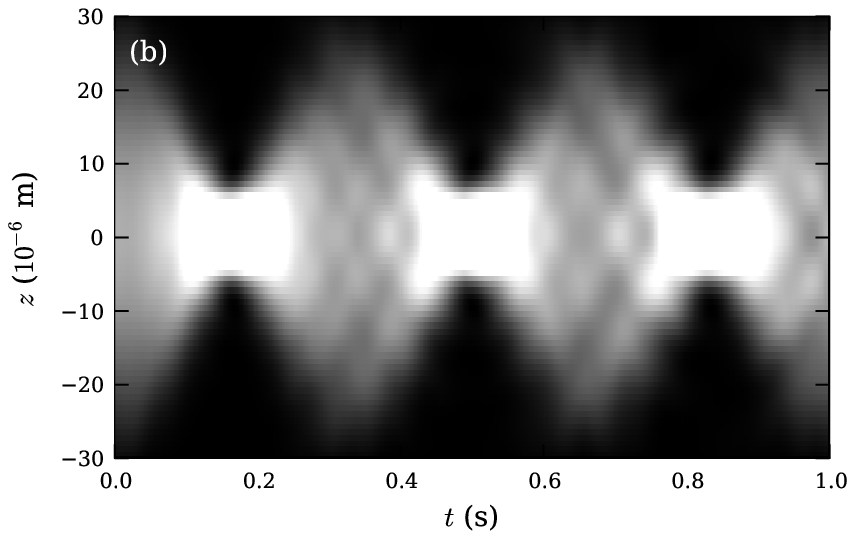}}
	\end{center}
	\caption{
		Oscillations of the one-dimensional density $\left|f_2(z,t)\right|^2$ with frequency $2.91$~Hz evaluated from Eq.~\ref{eq:effective-ho-solution}
		\subref{fig:density-analytic} and the linear density with frequency $3.00$~Hz simulated with the coupled 3D GPE (Eqs.~\ref{eq:cgpe})~\subref{fig:density-gpe}.
	}
	\label{fig:simulations-vs-analytics}
\end{figure}

In this Section we test the accuracy of the predictions of the effective 1D treatment, by comparing the above results with the full numerical simulations of the 3D GPE including collisional losses, described by the following equations:
\begin{equation}
	\begin{split}
		i\hbar\frac{\partial\Psi_1}{\partial t} &= \Big[-\frac{\hbar^2\nabla^2}{2m} + V(\mathbf{r}) + U_{11}\lvert\Psi_1\rvert^2 + U_{12}\lvert\Psi_2\rvert^2\\
			&- i\Gamma_1\Big]\Psi_1,\\
		i\hbar\frac{\partial\Psi_2}{\partial t} &= \Big[-\frac{\hbar^2\nabla^2}{2m} + V(\mathbf{r}) + U_{12}\lvert\Psi_1\rvert^2 + U_{22}\lvert\Psi_2\rvert^2\\
			&- i\Gamma_2\Big]\Psi_2.
	\end{split}
	\label{eq:cgpe-with-losses}
\end{equation}
Here the loss rates of species $1$ and $2$ are $\Gamma_1=\frac{\hbar}2 \gamma_{12}\left|\Psi_2\right|^2$ and $\Gamma_2=\frac{\hbar}2 (\gamma_{12}\left|\Psi_1\right|^2 + \gamma_{22}\left|\Psi_2\right|^2)$, and $\gamma_{12}$ and $\gamma_{22}$ are the two-body loss coefficients~\cite{Mertes07}.
The three-body loss rate is negligible at our typical BEC densities.
The simulations are performed using a symmetric split-step Fourier method~\cite{Sinkin03}.
The main idea of this method is the integration of the GPE by the separate application of nonlinear and differential operators (the latter being applied in Fourier space).
The symmetric version of the method additionally applies a differential operator in two steps, separated by the application of a nonlinear operator in a middle step, allowing it to reach a local error of the third order in the time step.
Free expansion of the BEC is simulated when needed on a grid with increased size ($128\times256\times256$ for a free expansion compared to $128\times32\times32$ in a trap) with the trapping potential set to $0$.
We also find the optimal regimes for precision measurements of $a_{12}$ and the dependence on various parameters.
Unless otherwise specified we use the \Rb~parameters for the states $\ket{1}\equiv\ket{F=1,~m_F=-1}$ and $\ket{2}\equiv\ket{F=2,~m_F=+1}$ (${a_{11} = 100.4\,a_0}$, ${a_{12} = 98.006\,a_0}$
and ${a_{22} = 95.44\,a_0}$) and the trap frequencies ($98.23$,~$101.0$,~$11.507$)~Hz.

For $N=10^5$ atoms and a transfer of $2.4\%$ of the atoms to state $\ket{2}$ the evolution of the simulated linear density of component
$\ket{2}$ ($n_{2l}=\int\Psi_2^*\Psi_2\,dxdy$,~\figref{fig:density-gpe})
resembles that from the effective 1D treatment (\figref{fig:density-analytic}) but exhibits a few additional distinct features.
Firstly, the frequency of the collective oscillations is $3.00$~Hz.
Secondly, additional periodic variations of the linear density with frequency $18.3(3)$~Hz are clearly visible.
These represent the monopole compression mode, which is also excited due to atomic interactions being altered by the transfer of atoms to state $\ket{2}$.
Indeed, the frequency of these fast oscillations is consistent with the estimated value of the lowest monopole mode frequency~$\sqrt{5/2}f_z = 18.2$~Hz for very elongated traps~\cite{Stringari96}.

\begin{figure}[!]
	\centering
	\subfigure{\label{fig:gpe-a12}\includegraphics{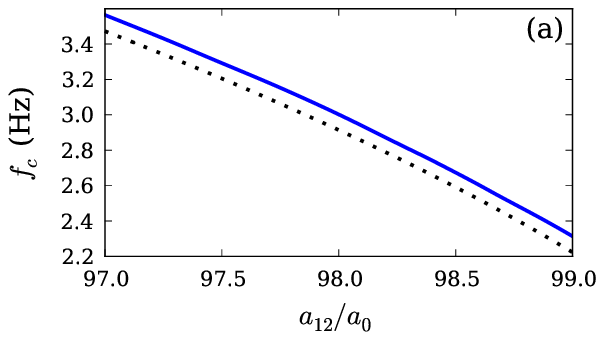}}\\
	\subfigure{\label{fig:gpe-a22}\includegraphics{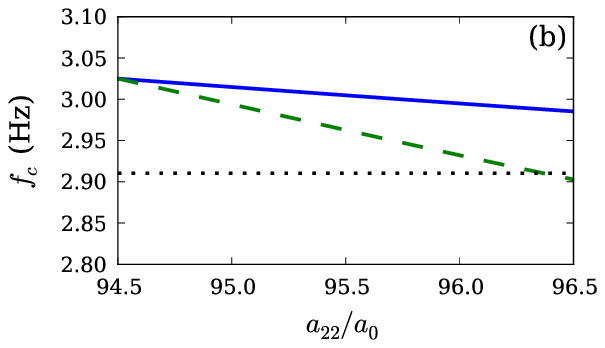}}\\
	\subfigure{\label{fig:gpe-N}\includegraphics{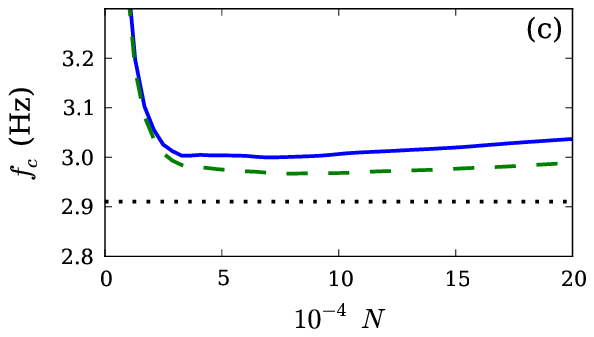}}
	\caption{
		\label{fig:simulations-frequencies}
		(Color online)
		Dependence of the collective oscillations frequency of component $\ket{2}$ on the scattering lengths $a_{12}$~\subref{fig:gpe-a12}
		and $a_{22}$~\subref{fig:gpe-a22} and on the total atom number $N$~\subref{fig:gpe-N}.
		Dotted lines are the analytical predictions of Eq.~\ref{eq:collective-oscillations-frequency}.
		The three-dimensional GPE simulations are represented by the solid lines ($\theta=\pi/10$) and dashed lines ($\theta=\pi/5$).
		$N = 10^5$ for~\subref{fig:gpe-a12} and~\subref{fig:gpe-a22}.
	}
\end{figure}

The frequency of the slow collective oscillations is very sensitive to variations in the value of $a_{12}$~(\figref{fig:gpe-a12})
and can be approximated in the range of interest by
\begin{equation}
	f_{\textrm{c}}\left(\xi_{12}\right) = \left(3.00 - 0.63\,\xi_{12} - 0.063\,\xi_{12}^2\right)~\textrm{Hz},
	\label{eq:fc_a12_poly}
\end{equation}
where $\xi_{12}=a_{12}/a_0 - 98.0$ and the preparation pulse area is $\theta=\pi/10$.
The simulated value is higher by $3\%$ than the value estimated from Eq.~\ref{eq:collective-oscillations-frequency} which does not account for
the dynamics of the BEC in the radial direction.
A one-dimensional treatment accounting for those dynamics can be derived~\cite{Kamchatnov04}; however the resulting equations can be solved only numerically.
The difference between Eq.~\ref{eq:collective-oscillations-frequency} and the GPE simulations depends weakly on $N$
which appears in our simulations as the derivative $\partial f_{\textrm{c}}/\partial N\sim 2\times 10^{-7}$.
The dependence on $a_{22}$ is greatly suppressed for small atom number in state $\ket{2}$~(\figref{fig:gpe-a22}).
For a mixture of two components prepared by a $\pi/10$-pulse, $f_{\textrm{c}}$ can be estimated from
\begin{equation}
	f_{\textrm{c}}\left(\xi_{22}\right) = \left(3.00 - 0.020\,\xi_{22}\right)~\textrm{Hz},
	\label{eq:fc_a22_poly}
\end{equation}
where $\xi_{22}=a_{22}/a_0 - 95.5$.
Sensitivity to the total atom number $N$ is also suppressed for small pulse areas $\theta$~(\figref{fig:gpe-N})
and sufficiently large atom numbers~(Eq.~\ref{eq:large-N-criterion}, \figref{fig:gpe-N}).
We have checked that the dependence of the frequency on the pulse area is not pronounced when $\theta\ll\pi/2$.
We use the first order derivatives of the collective oscillation frequencies on the experimental parameters in our error analysis later in the paper.
The simulations showed that for sufficiently large $N$ and small $\theta$, the collective oscillations frequency can be
used for precision measurements of $a_{12}/a_{11}$.
We have estimated the accuracy of measuring $a_{12}$ from the trap frequency $f_z$.
While $f_{\textrm{c}}$ is independent of $f_{\textrm{r}}$, Eqs.~\ref{eq:collective-oscillations-frequency} and~\ref{eq:fc_a12_poly} assume that
an axial trap frequency measurement with a precision ${\delta f_z/f_z=6\times10^{-4}}$ (\figref{fig:trap-frequency}) leads to an additional uncertainty in the $a_{12}$ measurement
of $\delta a_{12}=0.003\,a_0$.

\section{Experimental setup}
\label{sec:experimental-setup}

We generate an almost pure condensate of \Rb~atoms in state $\ket{1}$ in a cigar-shaped magnetic trap on an atom chip~\cite{Hall06}.
A two-photon microwave-radiofrequency (MW-RF) transition~\cite{Mertes07, Anderson09} is used for the fast transfer of a variable number of atoms
from state $\ket{1}$ to state $\ket{2}$.
The MW radiation ($f_{\textrm{MW}}\approx 6.8$~GHz) is applied by a half-wave dipole antenna located outside of the vacuum chamber and
the RF field ($f_{\textrm{RF}} \approx 3.2$~MHz) is coupled using two side wires on the chip~\cite{Hall06}.
During the transfer the MW field is red-detuned from the intermediate state $\ket{F=2, m_F=0}$ by $1$~MHz.
The two-photon Rabi frequency of the effective two-level system is $\Omega_{12}/(2\pi) = 500$~Hz and the two-photon detuning is $\Delta/(2\pi)=6.8$~Hz.

We use MW spectroscopy of the transition $\ket{1}\rightarrow\ket{F=2,m_F=0}$ to adjust the magnetic field at the trap bottom to the value $3.228(5)$~G
at which the first order Zeeman shift between states $\ket{1}$ and $\ket{2}$ is cancelled so that the atoms in both states experience almost the same
trapping potential~\cite{Harber02}.
For accurate knowledge of the total atom number $N$ we detect the atoms in both states in the same experimental realization.
At the end of the cycle, after the cloud is released from the trap, we employ adiabatic passage with MW radiation~\cite{Anderson09, Egorov11}
to transfer the state $\ket{1}$ atoms to the state $\ket{3}\equiv\ket{F=2,m_F=-1}$ with an efficiency of $98\%$.
During the time-of-flight fall the atoms in states $\ket{2}$ and $\ket{3}$ are spatially separated by a magnetic field gradient generated with a pulse
of current through the Z-wire on the chip.
A single absorption image of the atoms in the two states is taken using a $100~\upmu\textrm{s}$ pulse of probing light resonant with the
$F=2 \rightarrow F^{\prime}=3$ cyclic transition.
The optical resolution of our imaging system is estimated to be around $6~\upmu\textrm{m}$.
The experimental images are post-processed by a fringe-removal ``eigenface'' algorithm~\cite{Ockeloen10} to improve the signal-to-noise ratio.

\begin{figure}
	\begin{center}
		\includegraphics{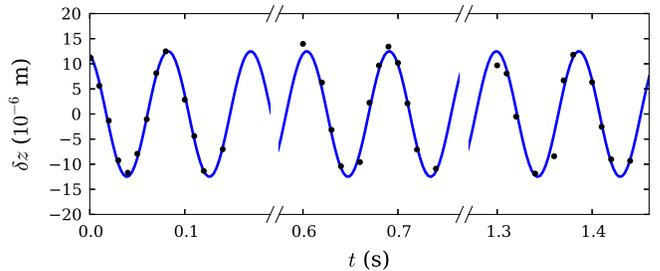}
	\end{center}
	\caption{
		(Color online)
		Dipole oscillations of state $\ket{1}$ BEC along the axial direction of the magnetic trap.
		The measured value of the axial trap frequency is $f_z=11.507(7)$~Hz.
	}
	\label{fig:trap-frequency}
\end{figure}
Accurate knowledge of the harmonic trap frequencies (especially $f_z$) is essential for precision measurements of the $a_{12}$ value.
We employ a standard method of dipole oscillations by suddenly shifting the trap along the measured axis, returning to the original position after half of a cycle and monitoring the periodic oscillations of the BEC in state $\ket{1}$~(\figref{fig:trap-frequency}).
The dipole oscillations in the axial direction are not damped over a long period of time.
From these measurements we infer the three harmonic oscillator frequencies of our trap:
$f_z = 11.507(7)$~Hz, $f_x = 98.23(5)$~Hz and $f_y = 101.0(5)$~Hz.
We checked numerically that anharmonicity of the trapping potentials on the atom chip does not affect the results.

\section{Measurement sequence and convergence of analysis}
\label{sec:convergence}

\begin{figure}
	\centering
	\subfigure{\label{fig:converge-a12}\includegraphics[width=0.48\columnwidth]{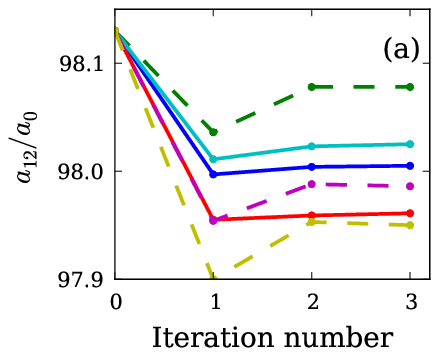}}
	\subfigure{\label{fig:converge-a22}\includegraphics[width=0.48\columnwidth]{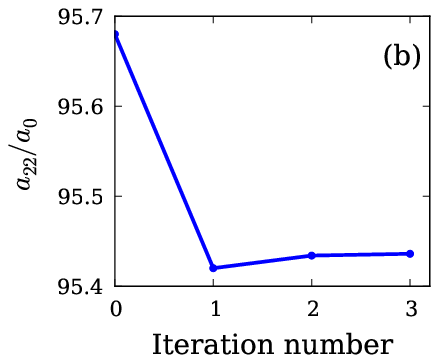}}
	\caption{\label{fig:convergence}
		(Color online)
		Iterative convergence of values of the interspecies scattering length $a_{12}$~\subref{fig:converge-a12}
		and the intraspecies scattering length $a_{22}$~\subref{fig:converge-a22}.
		Solid lines represent the results obtained with a preparation pulse area $\theta=\pi/10$, dashed lines are for $\theta=\pi/5$ in~\subref{fig:converge-a12}.
		Different colors in~\subref{fig:converge-a12} represent the data for different sets of measurements of $a_{12}$.
	}
\end{figure}

The time dependence of the collective oscillations of the lightly populated component $\ket{2}$ is crucially dependent on the $a_{12}/a_{11}$ ratio.
In order to account for the residual dependence on the scattering length $a_{22}$, the total atom number $N$ and the two-body loss coefficients,
we carry out measurements and analysis of the data in the following way.
We use the theoretical predictions of the scattering length values ($a_{12}=98.13\,a_0$ and $a_{22}=95.68\,a_0$)~\cite{Verhaar09} and
experimental values of the two-body loss coefficients measured at $8.32$~G
($\gamma_{12}=7.80\times10^{-20}~\textrm{m}^3/\textrm{s}$ and $\gamma_{22}=1.194\times10^{-19}~\textrm{m}^3/\textrm{s}$)~\cite{Mertes07}
as initial parameters.
In the first iteration we use these values in the interferometric calibration of the total atom number $N$~\cite{Egorov11} which we find to be consistent with the calibration using the condensation temperature.
We measure the two-body loss coefficients using the atom number calibration results.
Then we find a new value of $a_{12}$ from collective oscillations dynamics.
In the next step we find $a_{22}$ from Ramsey interferometry measurements with $\pi/10$ and $\pi/2$ preparation pulses.
At the end of the first iteration we find new values of $N$, $\gamma_{12}$, $\gamma_{22}$, $a_{12}$ and $a_{22}$ and
cycle through the same sequence of analysis several times until all values converge.
We find that $3$ iterations are sufficient for convergence~(\figref{fig:convergence}).

\section{Two-body loss coefficients}
\label{sec:loss-coefficients}

In the mean field approximation atom losses of a two-component BEC are described by the equations~\cite{Tojo09}
\begin{equation}
	\begin{split}
		\frac{dn_1}{dt} &= - \gamma_{12} n_1 n_2,\\
		\frac{dn_2}{dt} &= -\gamma_{22} n_2^2 - \gamma_{12} n_1 n_2,
		\label{eq:2c-losses}
	\end{split}
\end{equation}
where $n_1$ and $n_2$ are the densities of each BEC component, and $\gamma_{12}$ and $\gamma_{22}$ are two-body loss coefficients.
During an inelastic collision two condensed atoms in state $\ket{F=2,m_F=1}$ change their spin states to $\ket{F=2,m_F=0}$ and $\ket{F=2,m_F=2}$.
An atom in $\ket{F=2,m_F=0}$ state is lost from the magnetic trap and an atom in $\ket{F=2,m_F=2}$ acquires a potential energy of $624$~nK due to the gravitational sag and moves out of the trap.
The atoms in these states do not contribute to losses of the two-component BEC as they do not overlap with the condensed atoms.
Two atoms in states $\ket{F=2,m_F=1}$ and $\ket{F=1,m_F=-1}$ can spin flip to untrappable states $\ket{F=2,m_F=0}$ and $\ket{F=1,m_F=0}$ and do not further contribute to atom losses.
Here we neglect inelastic collisions with the background gas and three-body losses which do not contribute on a timescale of less than one second.

\begin{figure}
	\centering
	\vspace{3mm}
	\subfigure{\label{fig:gamma22}\includegraphics{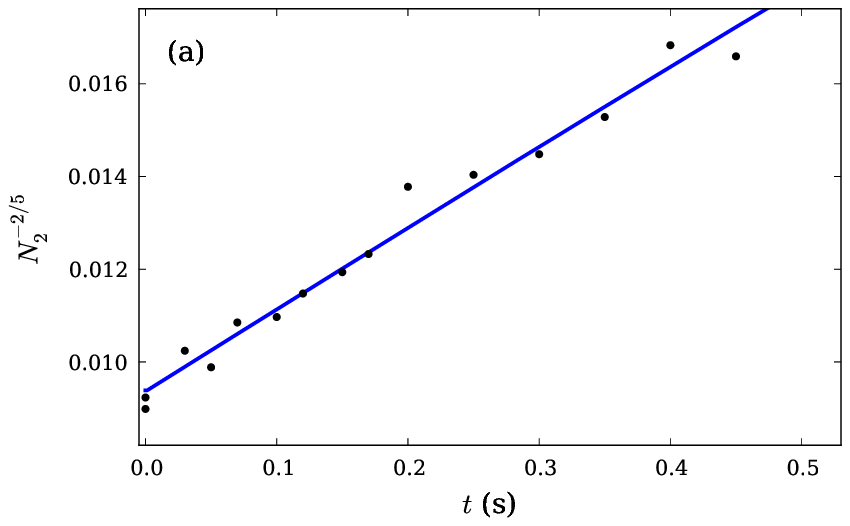}}\\
	\subfigure{\label{fig:gamma12}\includegraphics{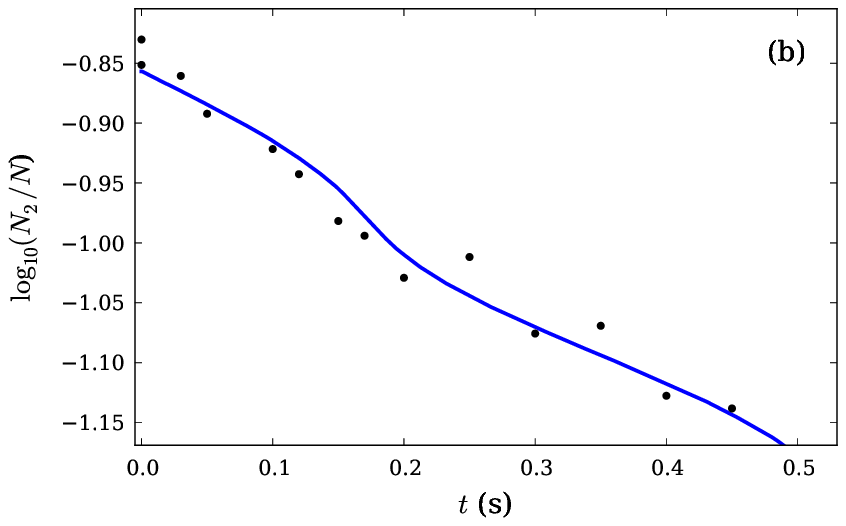}}
	\caption{\label{fig:loss-terms}
		(Color online)
		Measurement of intraspecies $\gamma_{22}$~\subref{fig:gamma22} and interspecies $\gamma_{12}$~\subref{fig:gamma12} two-body loss coefficients.
		The plot of $N^{-2/5}$ versus hold time $t$ is almost linear in the $\gamma_{22}$ measurement.
		The measurement of $\gamma_{12}$ is plotted on a semi-log scale, the increase in loss rate is observed at the point of maximum density of component $2$ ($0.17$~s) due to the influence of $\gamma_{22}$.
		Black points are experimental results, blue solid lines represent fits with GPE simulations.
	}
\end{figure}

In order to measure the $\gamma_{22}$ coefficient we produce a BEC in state $\ket{1}$ and then prepare a pure state $\ket{2}$ condensate with a $\pi$ pulse.
After a variable evolution time $t$, we release the condensate and measure the remaining number of atoms in state $\ket{2}$.
We apply a magnetic field gradient to separate in free fall the atoms in $\ket{2}$ and $\ket{F=2,m_F=2}$ states and only measure the population of state $\ket{2}$.
If the BEC adiabatically follows a TF profile during the lossy evolution, the loss of a single component is described for short times by~\cite{Tojo09}
\begin{equation}
	N_2^{-\frac25}(t) =
		N_2^{-\frac25}(0) +
		\left[\frac25 \frac {(2\pi)^{\frac15} 15^{\frac25}} {7a_{22}^{\frac35}}
		\left(
			\frac {m\bar{f}} {\hbar}
		\right)^{\frac65}
		\gamma_{22}\right] t,
	\label{eq:analytic-losses}
\end{equation}
where $\bar{f}=48.5$~Hz is the mean trap frequency.
Thus the value of $\gamma_{22}$ is given by the slope of the dependence $N_2^{-2/5}(t)$~(\figref{fig:gamma22}, solid line).
In order for the BEC to follow the trapping potential adiabatically during the loss process, the characteristic loss rate $\Gamma_2$ should be less than the trapping frequency $f_z$.
However, in our experiments these quantities are comparable, and we find the atom loss by fitting the GPE equations (Eqs.~\ref{eq:cgpe-with-losses}).
However, the results of the GPE simulations are very close to the linear dependence of Eq.~\ref{eq:analytic-losses} and are not distinguishable in~\figref{fig:gamma22}.

For the $\gamma_{12}$ measurements we prepare a superposition of two states with a $\pi/5$ pulse ($n_2\ll n_1$) and measure the remaining atom numbers in the two components after an evolution time $t$ (\figref{fig:gamma12}).
Under such conditions the loss process depends mostly on $\gamma_{12}$ rather than $\gamma_{22}$.
This allows us to slightly decouple the measurements from the value of $\gamma_{22}$.
We find $\gamma_{12}$ by fitting the experimental data with the coupled GPE simulations (Eqs.~\ref{eq:cgpe-with-losses}) using the iterated atom number $N$.

The exact values of the loss coefficients are strongly dependent on the initial total number of atoms~(Eq.~\ref{eq:analytic-losses}).
Using the converging sequence of Section~\ref{sec:convergence} we determine the values $\gamma_{12}=1.51(18)\times10^{-20}~\textrm{m}^3/\textrm{s}$ and $\gamma_{22}=8.1(3)\times10^{-20}~\textrm{m}^3/\textrm{s}$.
These values differ from the previously reported values of $\gamma_{22} = 10.4(10)\times10^{-20}~\textrm{m}^3/\textrm{s}$ measured at a magnetic field of $3.0$~G in a dipole trap~\cite{Tojo09} and $\gamma_{22}=11.94(19)\times10^{-20}~\textrm{m}^3/\textrm{s}$ and $\gamma_{12}=7.80(19)\times10^{-20}~\textrm{m}^3/\textrm{s}$ measured at $8.32$ G in a TOP magnetic trap~\cite{Mertes07}.
In a separate measurement of the evolution of cold non-condensed atoms prepared with a $\pi$ pulse in state $\ket{2}$ we determine $\gamma_{22}^{\prime}=16.9\times10^{-20}~\textrm{m}^3/\textrm{s}$ which is by a factor of two larger than our measured value of $\gamma_{22}$.
The two-particle correlation coefficient for noncondensed bosons is equal $2$~\cite{Harber02}, so our measurements of losses of condensed and noncondensed atoms are consistent with each other.

\section{Measurement of $a_{12}$ scattering length}
\label{sec:a12-measurements}

\begin{figure}
	\begin{center}
		\includegraphics[width=\columnwidth]{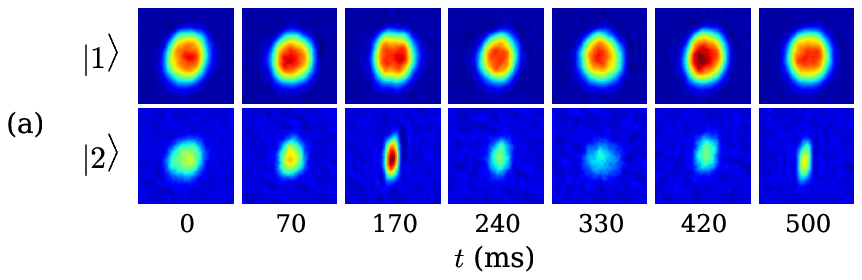}\\
		\quad
		\includegraphics{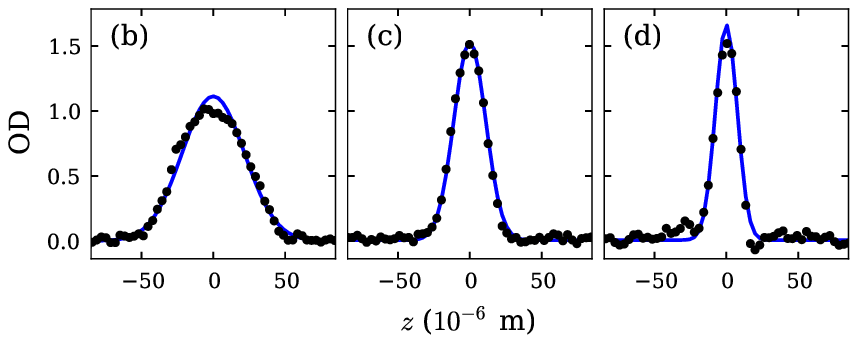}
	\end{center}
	\caption{
		(Color online)
		(a) Absorption images of two BEC components in the time-of-flight expansion for different evolution times after a $\pi/5$ preparation pulse.
		(b-d)~Central cross sections of the column density of component $2$ at evolution times of $0$~(b), $100$~(c) and $170$~(d)~ms.
		Black dots represent the measured optical density from the CCD pixels.
		Blue solid lines are the central cuts of the 2D Gaussian functions with axial widths of $21.7$~(b), $11.1$~(c) and $7.6$~(d)~$\upmu$m. 
	}
	\label{fig:density-profiles}
\end{figure}

We carry out six sets of measurements: three with $\pi/10$ pulses and three with $\pi/5$ pulses.
Oscillations of the axial width of component $2$ are excited (Sec.~\ref{sec:gpe-simulations}) and we image the column densities of both components in time-of-flight after various evolution times (Fig.~\ref{fig:density-profiles}).
The experiments are performed with two different times of free expansion ($6.6$ and $20.1$~ms) which are also included in the GPE simulations.
The period of the collective oscillations depends on the ratio $a_{12}/a_{11}$ rather than just the value of $a_{12}$ (Eq.~\ref{eq:collective-oscillations-frequency}) and we use the established value $a_{11}=100.40\,a_0$~\cite{Widera06,Mertes07}.

The two dimensional distribution of the column density of the second component is fitted with a 2D Gaussian function and from here we extract axial widths of the column density profiles.
The axial cross sections in the centre of the 2D profiles are shown in Figs.~\ref{fig:density-profiles}(b-d).
The choice of a Gaussian function to fit the experimental data originates from the fact that Eq.~\ref{eq:effective-ho-se} assumes that the ground state of component $2$ has a Gaussian shape.
Even when the initial BEC density profile has the shape given by the TF approximation, a Gaussian function fits the experimental cross section well (Fig.~\ref{fig:density-profiles}b) and is a good measure of the BEC width.
In our analysis we do not use second moments of the column density for width measurements because in this case the extracted BEC width has error bars larger by a factor of $\sim 20$.

\begin{figure}
	\centering
	\subfigure{\label{fig:width-experiment-pi10-nodrop}\includegraphics{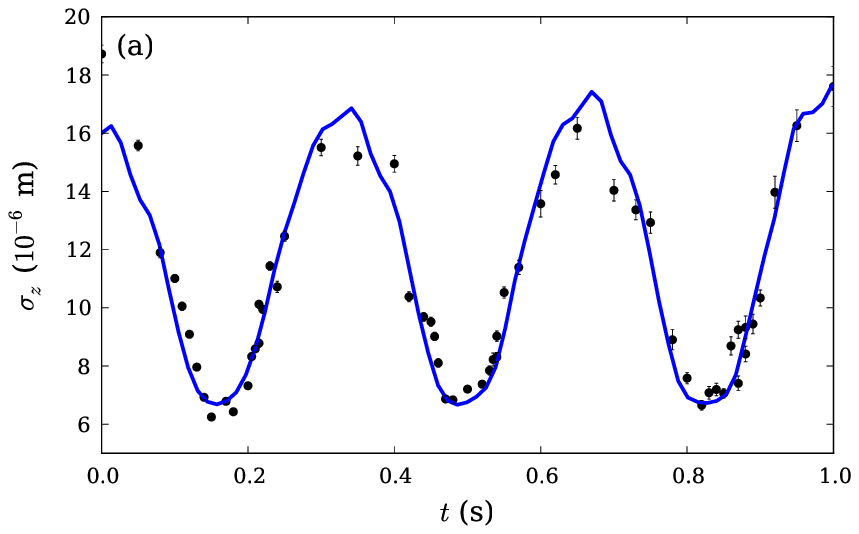}}
	\subfigure{\label{fig:width-experiment-pi5-nodrop}\includegraphics{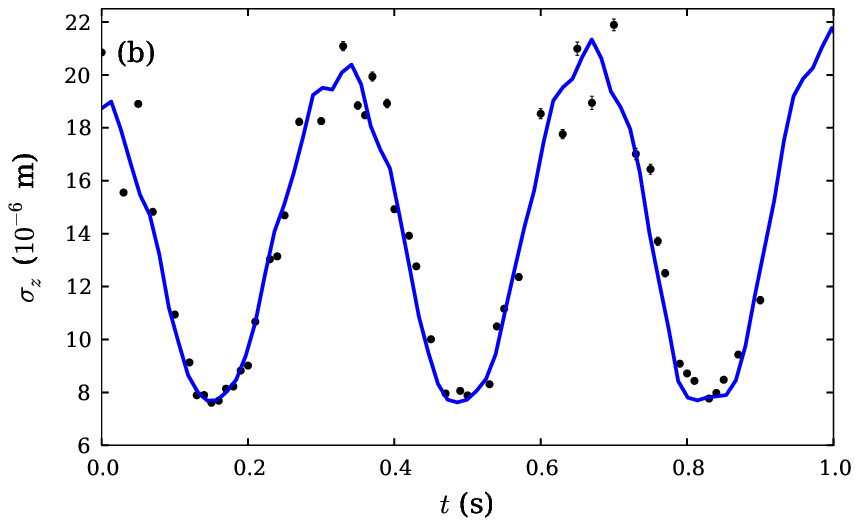}}

	\caption{\label{fig:focusing-dynamics}
		(Color online)
		Temporal evolution of the axial width of component $2$ in a superposition of two states prepared with a $\pi/10$~\subref{fig:width-experiment-pi10-nodrop} or $\pi/5$~\subref{fig:width-experiment-pi5-nodrop} pulse.
		The expansion times are $6.6$~ms~(a) and $20.1$~ms~(b), and the initial total atom numbers are $N=1.1\times 10^5$~(a) and $N=7.6\times 10^4$~(b).
		Black dots are the data extracted with the 2D Gaussian functions and error bars represent statistical errors of the fits.
		Blue solid lines are the results of the GPE simulations (Eq.~\ref{eq:cgpe-with-losses}) with $a_{12}=98.025\,a_0$~(a) and $97.986\,a_0$~(b), and $a_{22}=95.44\,a_0$.
	}
\end{figure}

Oscillations of the axial width of component $2$ with time are shown on Figs.~\ref{fig:width-experiment-pi10-nodrop} and~\ref{fig:width-experiment-pi5-nodrop} for $\pi/10$ and $\pi/5$ preparation pulses and a time-of-flight of $6.6$~ms.
Each point represents one experimental realization and error bars are the statistical uncertainty of the Gaussian fits of the recorded 2D column densities.
We fit the temporal dependence of the axial width with the coupled GPE equations (Eq.~\ref{eq:cgpe-with-losses}) varying $a_{12}$, $a_{22}$ and $N$ and using the iteration procedure of Sec.~\ref{sec:convergence}.
The extracted values of the scattering length $a_{12}$ for the six measurement sets are shown in the Table~\ref{tab:a12-results}.
\begin{table}
	\begin{tabular}{|c|cccccc|}
		\hline
		$\frac{a_{12}}{a_0}$ &
		$\frac{\delta_{\textrm{f}}a_{12}}{a_0}$ &
		$\frac{\partial f_{\textrm{c}}} {\partial N}$,~Hz &
		$\frac{\partial f_{\textrm{c}}} {\partial \theta},~\frac{\textrm{Hz}}{\textrm{rad}}$ &
		$\frac{\partial f_{\textrm{c}}} {\partial a_{22}},~\frac{\textrm{Hz}}{a_0}$ &
		$\frac{\partial f_{\textrm{c}}} {\partial a_{12}},~\frac{\textrm{Hz}}{a_0}$ &
		$\frac{\delta a_{12}}{a_0}$ \\
		\hline
		$98.005$ & $0.028$ & $1.39\times10^{-7}$ & $0.084$ & $0.020$ & $0.63$ & $0.037$ \\
		$97.961$ & $0.032$ & $2.8\times10^{-7}$ & $0.084$ & $0.020$ & $0.63$ & $0.046$ \\
		$98.025$ & $0.019$ & $2.8\times10^{-7}$ & $0.084$ & $0.020$ & $0.63$ & $0.033$ \\
		$98.078$ & $0.018$ & $1.57\times10^{-7}$ & $0.148$ & $0.062$ & $0.63$ & $0.042$ \\
		$97.950$ & $0.022$ & $2.1\times10^{-7}$ & $0.148$ & $0.062$ & $0.63$ & $0.050$ \\
		$97.986$ & $0.015$ & $2.1\times10^{-7}$ & $0.148$ & $0.062$ & $0.63$ & $0.042$ \\
		\hline
		\multicolumn{7}{|r|}{Weighted mean: $a_{12}=(98.006\pm0.016)\,a_0$}\\
		\hline
	\end{tabular}
	\caption{\label{tab:a12-results}
		Values of the scattering length $a_{12}$ extracted from the six experimental sets with uncertainties coming from different sources.
	}
\end{table}
Systematic errors are calculated for each set of measurements assuming $10\%$ uncertainty in the preparation pulse area $\theta$ and $15\%$ preparation noise in the initial total atom number $N$.
The combined uncertainty of $a_{12}$ consists of the fit error $\delta_{\textrm{f}} a_{12}$ and the systematics calculated from the slopes of the corresponding dependencies (Fig.~\ref{fig:simulations-frequencies})
\begin{equation}
	\begin{split}
	\delta a_{12} & =
		\left(
			\frac{\partial f_{\textrm{c}}} {\partial a_{12}}
		\right)^{-1}
		\Bigg(
			0.10\,\theta \frac{\partial f_{\textrm{c}}} {\partial \theta} +
			0.15\,N \frac{\partial f_{\textrm{c}}} {\partial N} \\
			& + \delta a_{22} \frac{\partial f_{\textrm{c}}} {\partial a_{22}}
		\Bigg) +
		\delta_{\textrm{f}} a_{12}.
	\end{split}
\end{equation}
There is a larger deviation of the calculated axial width from data for the zero evolution time compared with other data points.
However, this generates a small contribution to the uncertainty of the fit error $\delta_{f} a_{12}$.

The final value of $a_{12}=(98.006\pm 0.016)\,a_0$ is calculated as the weighted mean of the six results, where the weight coefficients are obtained from the errors of individual measurements (Table~\ref{tab:a12-results}).
The inclusion of quantum noise~\cite{Opanchuk11} and finite temperature effects~\cite{Sinatra11} can improve the precision of the measurements and will be the subject of further studies.

\section{Measurement of $a_{22}$ scattering length}
\label{sec:a22-measurement}

\begin{figure}
	\centering
	\hspace{10mm}\includegraphics[width=0.8\columnwidth]{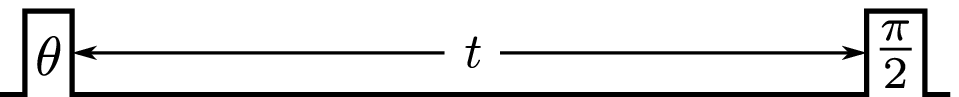}\\
	\subfigure{\label{fig:dual-ramsey-bec}\includegraphics{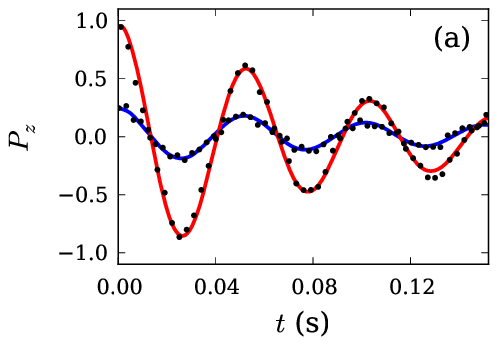}}
	\subfigure{\label{fig:thermal-ramsey}\includegraphics{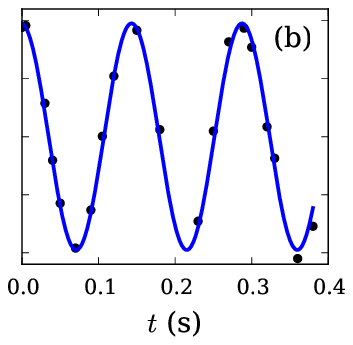}}
	\caption{\label{fig:dual-ramsey}
		(Color online)
		Ramsey interferometry with a variable area $\theta$ for the preparation pulse is used in measurements of the scattering length $a_{22}$.
		\subref{fig:dual-ramsey-bec} The normalized number difference $P_z$ oscillates with the evolution time $t$.
		Dots~-- data points, blue solid line~-- fitted GPE simulations for $\theta = \pi/10$, red solid line~-- fitted GPE simulations for $\theta = \pi/2$.
		\subref{fig:thermal-ramsey} Two-photon detuning $\Delta/(2\pi) = 6.82(10)$~Hz was measured in Ramsey interferometry of non-condensed atoms.
		Dots~-- data points, solid line~-- sinusoidal fit.
	}
\end{figure}

In this series of measurements we perform Ramsey interferometry of a BEC initially prepared in state $\ket{1}$ using a $\theta=\pi/2$ (or $\theta = \pi/10$) preparation pulse and a $\pi/2$ interrogation pulse (\figref{fig:dual-ramsey} top).
Our measured quantity, the normalized number difference $P_z=(N_2-N_1)/(N_2+N_1)$ (Figs.~\ref{fig:dual-ramsey-bec} and \subref{fig:thermal-ramsey}), where $N_1$ and $N_2$ are the populations of the two states after the interrogation pulse, oscillates with frequency determined by the detuning of the MW/RF radiation from the two-photon resonance and the collisional shift.
Two experimental sequences are intermixed in time in order to minimize effects of long-term drifts in the atom number $N$ and the MW frequency.
For the detuning measurement we perform Ramsey interferometry (two $\pi/2$ pulses) with a non-condensed cold atomic ensemble ($N\approx 1.8\times 10^4$, $T\approx 120$~nK, \figref{fig:thermal-ramsey}) in the same magnetic trap and find the two-photon detuning $\Delta=6.82(10)$~Hz.
For the cold atoms the collisional shift is estimated to be $\sim0.1$~Hz.
There is a damping of Ramsey interference fringes~(\figref{fig:dual-ramsey-bec}) driven by two factors: the spatially non-uniform growth of the relative phase (and the corresponding dephasing of the condensate wavefunctions) and asymmetric loss of the populations of two states (the state $\ket{2}$ has larger atom losses)~\cite{Anderson09, Egorov11}.

The collisional shift of dense condensed atoms depends not only on the particular values of the scattering lengths $a_{11}$, $a_{12}$ and $a_{22}$ but also on the total number of atoms $N$.
We use two different values of the preparation pulse area $\theta$ for the following reason.
If we assume that the atom number density $n(r)$ does not change with time the collisional shift is proportional to $n(r)(a_{11}-a_{22})$ for $\theta = \pi/2$~\cite{Harber02}.
For a $\theta = \pi/10$ preparation pulse the interaction-induced shift is proportional to $n(r)(a_{11}-a_{12})$.
Thus the ratio of the two shifts is independent of the atom number density and, therefore, independent of $N$.
In order to obtain a precise value of $a_{22}$, we run the GPE simulations with iterative values of the scattering lengths (Sect.~\ref{sec:convergence}) and fit the Ramsey fringe of $\theta = \pi/10$ (\figref{fig:dual-ramsey-bec}, solid line), keeping $N$ as a free parameter and taking into account that the fringe frequency is largely decoupled from the $a_{22}$ value.
Using the fitted value of $N$ we fit the $\theta=\pi/2$ data with the GPE simulations with the single variable parameter $a_{22}$ (\figref{fig:dual-ramsey-bec}, dashed line).
After convergence of the analysis of all measurements the value of the intraspecies scattering length for atoms in state $\ket{2}$ is found to be $a_{22} = 95.44(7)\,a_0$.

\section{Discussion}
\label{sec:discussion}

A number of different approaches have been used to evaluate the $s$-wave scattering lengths of \Rb~atoms prepared in the two lowest hyperfine states.
Here we restrict our discussion to the modeling and measurements of the scattering lengths $a_{11}$, $a_{22}$ and $a_{12}$ at reported values of the magnetic field.

An early calculation of the value $a_{11} = 106(6)\,a_0$~\cite{Vogels97} was based on knowledge of the $s$-wave bound-state energies of $^{87}\textrm{Rb}_2$.
Later, using the results of high-resolution molecular spectroscopy, the parameters of the atomic scattering potentials were calculated which allowed evaluation of $a_{11}=100.4(1)\,a_0$ and $a_{12} = 98.175\,a_0$~\cite{vanKempen02} at a magnetic field of $3.2$~G.
Cornell's group used the same parameters of the scattering potential to evaluate $a_{11}=100.44\,a_0$, $a_{12}=98.09\,a_0$ and $a_{22}=95.47\,a_0$~\cite{Harber02} for the same magnetic field.
Improved modeling of the atomic interaction potentials~\cite{Verhaar09} produced the latest theoretical values of the scattering lengths: $a_{11}=100.40\,a_0$, $a_{12} = 98.13\,a_0$ and $a_{22}=95.68\,a_0$~\cite{kokkelmans-private}.

Experiments with two-component BECs provided another way to measure the scattering properties of \Rb~atoms.
The transfer of the entire population from state $\ket{1}$ to state $\ket{2}$ produced a sudden change of collisional interactions in the condensate and triggered radial and axial collective oscillations of component $2$ in a time-orbiting-potential trap~\cite{Matthews98}.
GPE simulations were used to fit the temporal evolution of the radial and axial widths and yielded the ratio $a_{11}/a_{22} = 1.062(12)$.
If we assume $a_{11} = 100.40\,a_0$ then this measurement yields $a_{22} = 94.5(1.1)\,a_0$.
Collisional shift measurements~\cite{Harber02} in Ramsey interferometry of a two-component BEC and uncondensed atoms at the bias magnetic field of $3.23$~G produced a value for the difference of scattering lengths $a_{11}-a_{22} = 4.85(31)\,a_0$.
Collisionally driven spin oscillations of atom pairs trapped in an optical lattice at $0.24$~G allowed measurements of the scattering length differences in the $F = 1$ and $2$ manifolds~\cite{Widera06}.
The reported results consistently deviated by up to $10\%$ from the calculated values~\cite{vanKempen02}.
Use of the computed value $a_{11} = 100.40\,a_0$ and a fit of the observed oscillating ring-like structures in a two-component BEC at a bias magnetic field of $8.32$~G with the simulations of the coupled GPE equations have yielded $a_{12} = 97.66\,a_0$ and $a_{22} = 95.0\,a_0$~\cite{Mertes07}.
These three values of the scattering lengths are commonly used for modeling the Ramsey contrast and spin squeezing evolution in atom chip experiments on entangled atomic ensembles~\cite{Riedel10,Li09,Tacla10} carried out at a bias magnetic field of $3.23$~G or spin dynamics of 2CBEC in $\ket{F=1,m_F=1}$ and $\ket{F=2,m_F=-1}$ states~\cite{Tojo10,Niclas11}.

\begin{table}
	\centering
	\begin{tabular}{|c|cc|}
		\hline
		 & $a_{12}/a_0$ & $a_{22}/a_0$ \\
		\hline
		Matthews \textit{et al.} \cite{Matthews98} & & $94.5(1.1)$ \\
		van Kempen \textit{et al.} \cite{vanKempen02} & $98.175$ & \\
		Harber \textit{et al.} \cite{Harber02} & $98.09$ & $95.47$ \\
		Mertes \textit{et al.} \cite{Mertes07} & $97.66$ & $95.0$ \\
		Kokkelmans \cite{kokkelmans-private} & $98.13(10)$ & $95.68(10)$ \\
		\hline
		This work & $98.006(16)$ & $95.44(7)$ \\
		\hline
	\end{tabular}
	\caption{
		\label{tab:scattering-lengths}
		Calculated and measured values of $s$-wave scattering lengths in \Rb~for the magnetic field of $3.2$~G.
		Paper~\cite{Mertes07} reported the results for the magnetic field of $8.32$~G.
	}
\end{table}

In Table~\ref{tab:scattering-lengths} we compare the results of our measurements with previously reported values of $a_{12}$ and $a_{22}$.
Our value of $a_{12}$ is within $0.17\%$ of the theoretical values~\cite{vanKempen02,Harber02,kokkelmans-private}; however the statistical uncertainty of our measured value is ten times smaller.
All reported theoretical values are consistently shifted towards higher values.
Our value of $a_{12} = 98.006\,a_0$ deviates from that of~\cite{Mertes07} by $0.35\%$.
The dependence of $a_{12}$ on the magnetic field can explain this very significant deviation.
Our measured value of $a_{22}$ is very close to that of~\cite{Harber02} and is within $0.25\%$ of the latest theoretical evaluation~\cite{kokkelmans-private} but is well outside the statistical uncertainty $0.07\%$ of our measurement.

Precise knowledge of collisional loss rates is important as this limits the BEC
coherence time~\cite{Egorov11,Opanchuk11}, the effectiveness of spin-squeezing~\cite{Li08,Opanchuk11}
and contains important information about the properties of inter-atomic interaction potentials~\cite{vanKempen02}.
Two-body losses are usually the dominant source of losses in binary BECs.

In this work we have focused on collisions of \Rb~atoms in states $\ket{1}$ and $\ket{2}$ and the two-body loss
coefficients $\gamma_{12}$ and $\gamma_{22}$.
The coefficient $\gamma_{12}$ was initially calculated work by analyzing spectroscopic studies~\cite{vanKempen02}.
The reported value of the imaginary part of a complex scattering length $\Im(a(B))=-0.02\,a_0$ at the magnetic field of $3.23$~G can be used to evaluate the two-body loss rate by $\gamma_{12}=-4h/m\,\Im(a(B))=1.9\times 10^{-20}~\textrm{m}^3/\textrm{s}$.
Also this work predicted a weak Feshbach resonance at a magnetic field of $1.9$~G (where $\gamma_{12}$ increases by factor of $2$) which, however, has not yet been observed.
The two-body loss coefficients $\gamma_{12}=7.80\times10^{-20}~\textrm{m}^3/\textrm{s}$ and $\gamma_{22}=11.94\times10^{-20}~\textrm{m}^3/\textrm{s}$ were first experimentally characterized at the magnetic field of $8.32$~G~\cite{Mertes07}.
Another experimental measurement provided $\gamma_{22}=10.4(10)\times10^{-19}~\textrm{m}^3/\textrm{s}$ at a bias magnetic field of $3.0$~G assuming the two-body loss coefficient for state $\ket{F=2,m_F=-1}$ to be the same as for $\ket{F=2,m_F=+1}$ at low fields~\cite{Tojo09}.
The latter work carried out a comprehensive study of two-body losses of $F=2$ \Rb~atoms trapped in an optical trap.
All magnetic states were trapped and the presence of $m_F=0$ or $+2$ atoms may have influenced the reported results.

\begin{table}
	\centering
	\begin{tabular}{|c|cc|}
		\hline
		& $\gamma_{12} (\textrm{m}^3/\textrm{s})$ & $\gamma_{22} (\textrm{m}^3/\textrm{s})$ \\
		\hline
		van Kempen \textit{et al.} \cite{vanKempen02} & $1.9\times10^{-20}$ &  \\
		Mertes \textit{et al.} \cite{Mertes07} & $7.80(19)\times10^{-20}$ & $11.94(19)\times10^{-20}$\\
		Tojo \textit{et al.} \cite{Tojo09} & & $10.4(10)\times10^{-20}$ \\
		\hline
		This work & $1.51(18)\times 10^{-20}$ & $8.1(3)\times 10^{-20}$ \\
		\hline
	\end{tabular}
	\caption{
		\label{tab:loss-coefficients}
		Calculated and measured values of two-body loss coefficients in \Rb~for magnetic fields of $3.2$~G~\cite{vanKempen02}, $8.32$~G~\cite{Mertes07} and $3.0$~G~\cite{Tojo09}.
	}
\end{table}

In Table~\ref{tab:loss-coefficients} we compare the results of our measurements with previous theoretical and experimental investigations.
Our value of the interspecies two-body loss coefficient $\gamma_{12}$ is much closer to the theoretical result~\cite{vanKempen02} than the previous experimental result~\cite{Mertes07}.
Also our measurement of $\gamma_{22}$ is slightly smaller than the results obtained in other experimental works~\cite{Mertes07,Tojo09}.


\section{Conclusion}

We have presented a new technique for precision measurement of the interspecies scattering length $a_{12}$ in two-component Bose-Einstein condensates which employs collective oscillations of a less populated component.
The oscillations can be described either by simulations using coupled Gross-Pitaevskii equations or by the analytical model which we have developed.
Systematic errors such as uncertainty in the total number of atoms, the effect of one of the interspecies scattering lengths and imperfect preparation of the two-component mixture contribute very little to the uncertainty of the measured $a_{12}$.
We have applied the technique to measure the ratio of interspecies to intraspecies scattering lengths $a_{12}/a_{11}$ for states $\ket{1}\equiv\ket{F=1,m_F=-1}$ and $\ket{2}\equiv\ket{F=2,m_F=1}$ in \Rb~with an uncertainty of $0.016\%$.
Using a calculated value of $a_{11}=100.40\,a_0$, which is assumed to be established and error-free, we evaluated $a_{12}=98.006(16)\,a_0$.
The relative uncertainty of our measurement of $1.6\times10^{-4}$ is applied to the measured ratio of $a_{12}/a_{11} = 0.97616(16)$ and,
if the established value of $a_{11}$ changes in future measurements, this will proportionally affect our reported value of $a_{12}$.
We also have measured the intraspecies scattering length $a_{22}=95.44(7)\,a_0$ using Ramsey interferometry and the two-body loss coefficients $\gamma_{12}=1.51(18)\times 10^{-20}~\textrm{m}^3/\textrm{s}$ and $\gamma_{22}=8.1(3)\times 10^{-20}~\textrm{m}^3/\textrm{s}$ by fitting atom losses in the two-component BEC with the results of simulations of the coupled GPE.

Comparison of our results for $a_{12}$ and $a_{22}$ with the theoretical predictions and experimental measurements are presented in Tab.~\ref{tab:scattering-lengths}.
Our results show good agreement with recent theoretical calculations~\cite{kokkelmans-private}.
The residual deviations from the theoretical predictions could be caused either by uncertainty in the theoretical method or by unaccounted effects in our experiments.
The magnetic trapping potential on the atom chip is slightly anharmonic; however we have found that anharmonicity does not affect the GPE simulations.
Given the high precision of our measurements, the results could be affected by quantum dynamics beyond mean-field theory.
Another unaccounted contribution is from the finite temperature of the ensemble.
We will include quantum dynamics at finite temperatures into the simulations in future experiments~\cite{Opanchuk11}.

The present technique of $a_{12}$ measurement might be extremely useful in characterization of narrow interspecies Feshbach resonances, such as RF-induced Feshbach resonances~\cite{Kaufman09,Tscherbul10,Papoular10}.
Precision measurements of scattering lengths in the vicinity of narrow Feshbach resonances might also allow drifts of electron/proton mass ratio to be monitored~\cite{Chin06}.

We acknowledge helpful discussions with S.~Whitlock.
The project is supported by the ARC Centre of Excellence for Quantum-Atom Optics, and an ARC LIEF grant LE0668398.

\bibliography{article}

\begin{thebibliography}{36}%
\makeatletter
\providecommand \@ifxundefined [1]{%
 \@ifx{#1\undefined}
}%
\providecommand \@ifnum [1]{%
 \ifnum #1\expandafter \@firstoftwo
 \else \expandafter \@secondoftwo
 \fi
}%
\providecommand \@ifx [1]{%
 \ifx #1\expandafter \@firstoftwo
 \else \expandafter \@secondoftwo
 \fi
}%
\providecommand \natexlab [1]{#1}%
\providecommand \enquote  [1]{``#1''}%
\providecommand \bibnamefont  [1]{#1}%
\providecommand \bibfnamefont [1]{#1}%
\providecommand \citenamefont [1]{#1}%
\providecommand \href@noop [0]{\@secondoftwo}%
\providecommand \href [0]{\begingroup \@sanitize@url \@href}%
\providecommand \@href[1]{\@@startlink{#1}\@@href}%
\providecommand \@@href[1]{\endgroup#1\@@endlink}%
\providecommand \@sanitize@url [0]{\catcode `\\12\catcode `\$12\catcode
  `\&12\catcode `\#12\catcode `\^12\catcode `\_12\catcode `\%12\relax}%
\providecommand \@@startlink[1]{}%
\providecommand \@@endlink[0]{}%
\providecommand \url  [0]{\begingroup\@sanitize@url \@url }%
\providecommand \@url [1]{\endgroup\@href {#1}{\urlprefix }}%
\providecommand \urlprefix  [0]{URL }%
\providecommand \Eprint [0]{\href }%
\providecommand \doibase [0]{http://dx.doi.org/}%
\providecommand \selectlanguage [0]{\@gobble}%
\providecommand \bibinfo  [0]{\@secondoftwo}%
\providecommand \bibfield  [0]{\@secondoftwo}%
\providecommand \translation [1]{[#1]}%
\providecommand \BibitemOpen [0]{}%
\providecommand \bibitemStop [0]{}%
\providecommand \bibitemNoStop [0]{.\EOS\space}%
\providecommand \EOS [0]{\spacefactor3000\relax}%
\providecommand \BibitemShut  [1]{\csname bibitem#1\endcsname}%
\let\auto@bib@innerbib\@empty
\bibitem [{\citenamefont {Mertes}\ \emph {et~al.}(2007)\citenamefont {Mertes},
  \citenamefont {Merrill}, \citenamefont {Carretero-Gonz{\'{a}}lez},
  \citenamefont {Frantzeskakis}, \citenamefont {Kevrekidis},\ and\
  \citenamefont {Hall}}]{Mertes07}%
  \BibitemOpen
  \bibfield  {author} {\bibinfo {author} {\bibfnamefont {K.~M.}\ \bibnamefont
  {Mertes}}, \bibinfo {author} {\bibfnamefont {J.~W.}\ \bibnamefont {Merrill}},
  \bibinfo {author} {\bibfnamefont {R.}~\bibnamefont
  {Carretero-Gonz{\'{a}}lez}}, \bibinfo {author} {\bibfnamefont {D.~J.}\
  \bibnamefont {Frantzeskakis}}, \bibinfo {author} {\bibfnamefont {P.~G.}\
  \bibnamefont {Kevrekidis}}, \ and\ \bibinfo {author} {\bibfnamefont {D.~S.}\
  \bibnamefont {Hall}},\ }\href@noop {} {\bibfield  {journal} {\bibinfo
  {journal} {Phys. Rev. Lett.}\ }\textbf {\bibinfo {volume} {99}},\ \bibinfo
  {pages} {190402} (\bibinfo {year} {2007})}\BibitemShut {NoStop}%
\bibitem [{\citenamefont {Pitaevskii}\ and\ \citenamefont
  {Stringari}(2003)}]{Pitaevskii03}%
  \BibitemOpen
  \bibfield  {author} {\bibinfo {author} {\bibfnamefont {L.}~\bibnamefont
  {Pitaevskii}}\ and\ \bibinfo {author} {\bibfnamefont {S.}~\bibnamefont
  {Stringari}},\ }\href@noop {} {\emph {\bibinfo {title} {{Bose-Einstein
  Condensation}}}}\ (\bibinfo  {publisher} {{Oxford University Press}},\
  \bibinfo {year} {{2003}})\BibitemShut {NoStop}%
\bibitem [{\citenamefont {Weld}\ \emph {et~al.}(2010)\citenamefont {Weld},
  \citenamefont {Miyake}, \citenamefont {Medley}, \citenamefont {Pritchard},\
  and\ \citenamefont {Ketterle}}]{Weld10}%
  \BibitemOpen
  \bibfield  {author} {\bibinfo {author} {\bibfnamefont {D.~M.}\ \bibnamefont
  {Weld}}, \bibinfo {author} {\bibfnamefont {H.}~\bibnamefont {Miyake}},
  \bibinfo {author} {\bibfnamefont {P.}~\bibnamefont {Medley}}, \bibinfo
  {author} {\bibfnamefont {D.~E.}\ \bibnamefont {Pritchard}}, \ and\ \bibinfo
  {author} {\bibfnamefont {W.}~\bibnamefont {Ketterle}},\ }\href@noop {}
  {\bibfield  {journal} {\bibinfo  {journal} {Phys. Rev. A}\ }\textbf {\bibinfo
  {volume} {82}},\ \bibinfo {pages} {051603(R)} (\bibinfo {year}
  {2010})}\BibitemShut {NoStop}%
\bibitem [{\citenamefont {Verhaar}\ \emph {et~al.}(2009)\citenamefont
  {Verhaar}, \citenamefont {van Kempen},\ and\ \citenamefont
  {Kokkelmans}}]{Verhaar09}%
  \BibitemOpen
  \bibfield  {author} {\bibinfo {author} {\bibfnamefont {B.~J.}\ \bibnamefont
  {Verhaar}}, \bibinfo {author} {\bibfnamefont {E.~G.~M.}\ \bibnamefont {van
  Kempen}}, \ and\ \bibinfo {author} {\bibfnamefont {S.~J. J. M.~F.}\
  \bibnamefont {Kokkelmans}},\ }\href@noop {} {\bibfield  {journal} {\bibinfo
  {journal} {Phys. Rev. A}\ }\textbf {\bibinfo {volume} {79}},\ \bibinfo
  {pages} {032711} (\bibinfo {year} {2009})}\BibitemShut {NoStop}%
\bibitem [{\citenamefont {van Kempen}\ \emph {et~al.}(2002)\citenamefont {van
  Kempen}, \citenamefont {Kokkelmans}, \citenamefont {Heinzen},\ and\
  \citenamefont {Verhaar}}]{vanKempen02}%
  \BibitemOpen
  \bibfield  {author} {\bibinfo {author} {\bibfnamefont {E.~G.~M.}\
  \bibnamefont {van Kempen}}, \bibinfo {author} {\bibfnamefont {S.~J. J.
  M.~F.}\ \bibnamefont {Kokkelmans}}, \bibinfo {author} {\bibfnamefont {D.~J.}\
  \bibnamefont {Heinzen}}, \ and\ \bibinfo {author} {\bibfnamefont {B.~J.}\
  \bibnamefont {Verhaar}},\ }\href@noop {} {\bibfield  {journal} {\bibinfo
  {journal} {Phys. Rev. Lett.}\ }\textbf {\bibinfo {volume} {88}},\ \bibinfo
  {pages} {093201} (\bibinfo {year} {2002})}\BibitemShut {NoStop}%
\bibitem [{\citenamefont {Harber}\ \emph {et~al.}(2002)\citenamefont {Harber},
  \citenamefont {Lewandowski}, \citenamefont {McGuirk},\ and\ \citenamefont
  {Cornell}}]{Harber02}%
  \BibitemOpen
  \bibfield  {author} {\bibinfo {author} {\bibfnamefont {D.~M.}\ \bibnamefont
  {Harber}}, \bibinfo {author} {\bibfnamefont {H.~J.}\ \bibnamefont
  {Lewandowski}}, \bibinfo {author} {\bibfnamefont {J.~M.}\ \bibnamefont
  {McGuirk}}, \ and\ \bibinfo {author} {\bibfnamefont {E.~A.}\ \bibnamefont
  {Cornell}},\ }\href@noop {} {\bibfield  {journal} {\bibinfo  {journal} {Phys.
  Rev. A}\ }\textbf {\bibinfo {volume} {66}},\ \bibinfo {pages} {053616}
  (\bibinfo {year} {2002})}\BibitemShut {NoStop}%
\bibitem [{\citenamefont {Deutsch}\ \emph {et~al.}(2010)\citenamefont
  {Deutsch}, \citenamefont {Ramirez-Martinez}, \citenamefont {Lacro\^{u}te},
  \citenamefont {Reinhard}, \citenamefont {Schneider}, \citenamefont {Fuchs},
  \citenamefont {Pi\'{e}chon}, \citenamefont {Lalo\"{e}}, \citenamefont
  {Reichel},\ and\ \citenamefont {Rosenbusch}}]{Deutsch10}%
  \BibitemOpen
  \bibfield  {author} {\bibinfo {author} {\bibfnamefont {C.}~\bibnamefont
  {Deutsch}}, \bibinfo {author} {\bibfnamefont {F.}~\bibnamefont
  {Ramirez-Martinez}}, \bibinfo {author} {\bibfnamefont {C.}~\bibnamefont
  {Lacro\^{u}te}}, \bibinfo {author} {\bibfnamefont {F.}~\bibnamefont
  {Reinhard}}, \bibinfo {author} {\bibfnamefont {T.}~\bibnamefont {Schneider}},
  \bibinfo {author} {\bibfnamefont {J.~N.}\ \bibnamefont {Fuchs}}, \bibinfo
  {author} {\bibfnamefont {F.}~\bibnamefont {Pi\'{e}chon}}, \bibinfo {author}
  {\bibfnamefont {F.}~\bibnamefont {Lalo\"{e}}}, \bibinfo {author}
  {\bibfnamefont {J.}~\bibnamefont {Reichel}}, \ and\ \bibinfo {author}
  {\bibfnamefont {P.}~\bibnamefont {Rosenbusch}},\ }\href@noop {} {\bibfield
  {journal} {\bibinfo  {journal} {Phys. Rev. Lett.}\ }\textbf {\bibinfo
  {volume} {105}},\ \bibinfo {pages} {020401} (\bibinfo {year}
  {2010})}\BibitemShut {NoStop}%
\bibitem [{\citenamefont {Egorov}\ \emph {et~al.}(2011)\citenamefont {Egorov},
  \citenamefont {Anderson}, \citenamefont {Ivannikov}, \citenamefont
  {Opanchuk}, \citenamefont {Drummond}, \citenamefont {Hall},\ and\
  \citenamefont {Sidorov}}]{Egorov11}%
  \BibitemOpen
  \bibfield  {author} {\bibinfo {author} {\bibfnamefont {M.}~\bibnamefont
  {Egorov}}, \bibinfo {author} {\bibfnamefont {R.~P.}\ \bibnamefont
  {Anderson}}, \bibinfo {author} {\bibfnamefont {V.}~\bibnamefont {Ivannikov}},
  \bibinfo {author} {\bibfnamefont {B.}~\bibnamefont {Opanchuk}}, \bibinfo
  {author} {\bibfnamefont {P.}~\bibnamefont {Drummond}}, \bibinfo {author}
  {\bibfnamefont {B.~V.}\ \bibnamefont {Hall}}, \ and\ \bibinfo {author}
  {\bibfnamefont {A.~I.}\ \bibnamefont {Sidorov}},\ }\href@noop {} {\bibfield
  {journal} {\bibinfo  {journal} {Phys. Rev. A}\ }\textbf {\bibinfo {volume}
  {84}},\ \bibinfo {pages} {021605(R)} (\bibinfo {year} {2011})}\BibitemShut
  {NoStop}%
\bibitem [{\citenamefont {Riedel}\ \emph {et~al.}(2010)\citenamefont {Riedel},
  \citenamefont {B{\"{o}}hi}, \citenamefont {Li}, \citenamefont {H{\"{a}}nsch},
  \citenamefont {Sinatra},\ and\ \citenamefont {Treutlein}}]{Riedel10}%
  \BibitemOpen
  \bibfield  {author} {\bibinfo {author} {\bibfnamefont {M.~F.}\ \bibnamefont
  {Riedel}}, \bibinfo {author} {\bibfnamefont {P.}~\bibnamefont {B{\"{o}}hi}},
  \bibinfo {author} {\bibfnamefont {Y.}~\bibnamefont {Li}}, \bibinfo {author}
  {\bibfnamefont {T.~W.}\ \bibnamefont {H{\"{a}}nsch}}, \bibinfo {author}
  {\bibfnamefont {A.}~\bibnamefont {Sinatra}}, \ and\ \bibinfo {author}
  {\bibfnamefont {P.}~\bibnamefont {Treutlein}},\ }\href@noop {} {\bibfield
  {journal} {\bibinfo  {journal} {Nature}\ }\textbf {\bibinfo {volume} {464}},\
  \bibinfo {pages} {1170} (\bibinfo {year} {2010})}\BibitemShut {NoStop}%
\bibitem [{\citenamefont {Pu}\ and\ \citenamefont {Bigelow}(1998)}]{Pu98}%
  \BibitemOpen
  \bibfield  {author} {\bibinfo {author} {\bibfnamefont {H.}~\bibnamefont
  {Pu}}\ and\ \bibinfo {author} {\bibfnamefont {N.~P.}\ \bibnamefont
  {Bigelow}},\ }\href@noop {} {\bibfield  {journal} {\bibinfo  {journal} {Phys.
  Rev. Lett.}\ }\textbf {\bibinfo {volume} {80}},\ \bibinfo {pages} {1130}
  (\bibinfo {year} {1998})}\BibitemShut {NoStop}%
\bibitem [{\citenamefont {Tojo}\ \emph {et~al.}(2009)\citenamefont {Tojo},
  \citenamefont {Hayashi}, \citenamefont {Tanabe}, \citenamefont {Hirano},
  \citenamefont {Kawaguchi}, \citenamefont {Saito},\ and\ \citenamefont
  {Ueda}}]{Tojo09}%
  \BibitemOpen
  \bibfield  {author} {\bibinfo {author} {\bibfnamefont {S.}~\bibnamefont
  {Tojo}}, \bibinfo {author} {\bibfnamefont {T.}~\bibnamefont {Hayashi}},
  \bibinfo {author} {\bibfnamefont {T.}~\bibnamefont {Tanabe}}, \bibinfo
  {author} {\bibfnamefont {T.}~\bibnamefont {Hirano}}, \bibinfo {author}
  {\bibfnamefont {Y.}~\bibnamefont {Kawaguchi}}, \bibinfo {author}
  {\bibfnamefont {H.}~\bibnamefont {Saito}}, \ and\ \bibinfo {author}
  {\bibfnamefont {M.}~\bibnamefont {Ueda}},\ }\href@noop {} {\bibfield
  {journal} {\bibinfo  {journal} {Phys. Rev. A}\ }\textbf {\bibinfo {volume}
  {80}},\ \bibinfo {pages} {042704} (\bibinfo {year} {2009})}\BibitemShut
  {NoStop}%
\bibitem [{\citenamefont {Salasnich}\ \emph {et~al.}(2002)\citenamefont
  {Salasnich}, \citenamefont {Parola},\ and\ \citenamefont
  {Reatto}}]{Salasnich02}%
  \BibitemOpen
  \bibfield  {author} {\bibinfo {author} {\bibfnamefont {L.}~\bibnamefont
  {Salasnich}}, \bibinfo {author} {\bibfnamefont {A.}~\bibnamefont {Parola}}, \
  and\ \bibinfo {author} {\bibfnamefont {L.}~\bibnamefont {Reatto}},\
  }\href@noop {} {\bibfield  {journal} {\bibinfo  {journal} {Phys. Rev. A}\
  }\textbf {\bibinfo {volume} {65}},\ \bibinfo {pages} {043614} (\bibinfo
  {year} {2002})}\BibitemShut {NoStop}%
\bibitem [{\citenamefont {Massignan}\ and\ \citenamefont
  {Modugno}(2003)}]{Massignan03}%
  \BibitemOpen
  \bibfield  {author} {\bibinfo {author} {\bibfnamefont {P.}~\bibnamefont
  {Massignan}}\ and\ \bibinfo {author} {\bibfnamefont {M.}~\bibnamefont
  {Modugno}},\ }\href@noop {} {\bibfield  {journal} {\bibinfo  {journal} {Phys.
  Rev. A}\ }\textbf {\bibinfo {volume} {67}},\ \bibinfo {pages} {023614}
  (\bibinfo {year} {2003})}\BibitemShut {NoStop}%
\bibitem [{\citenamefont {Kamchatnov}\ and\ \citenamefont
  {Shchesnovich}(2004)}]{Kamchatnov04}%
  \BibitemOpen
  \bibfield  {author} {\bibinfo {author} {\bibfnamefont {A.~M.}\ \bibnamefont
  {Kamchatnov}}\ and\ \bibinfo {author} {\bibfnamefont {V.~S.}\ \bibnamefont
  {Shchesnovich}},\ }\href@noop {} {\bibfield  {journal} {\bibinfo  {journal}
  {Phys. Rev. A}\ }\textbf {\bibinfo {volume} {70}},\ \bibinfo {pages} {023604}
  (\bibinfo {year} {2004})}\BibitemShut {NoStop}%
\bibitem [{\citenamefont {Young-S.}\ \emph {et~al.}(2010)\citenamefont
  {Young-S.}, \citenamefont {Salasnich},\ and\ \citenamefont
  {Adhikari}}]{Young-S10}%
  \BibitemOpen
  \bibfield  {author} {\bibinfo {author} {\bibfnamefont {L.~E.}\ \bibnamefont
  {Young-S.}}, \bibinfo {author} {\bibfnamefont {L.}~\bibnamefont {Salasnich}},
  \ and\ \bibinfo {author} {\bibfnamefont {S.~K.}\ \bibnamefont {Adhikari}},\
  }\href@noop {} {\bibfield  {journal} {\bibinfo  {journal} {Phys. Rev. A}\
  }\textbf {\bibinfo {volume} {82}},\ \bibinfo {pages} {053601} (\bibinfo
  {year} {2010})}\BibitemShut {NoStop}%
\bibitem [{\citenamefont {P\'erez-Garc\'\i{}a}\ \emph
  {et~al.}(1996)\citenamefont {P\'erez-Garc\'\i{}a}, \citenamefont {Michinel},
  \citenamefont {Cirac}, \citenamefont {Lewenstein},\ and\ \citenamefont
  {Zoller}}]{Perez-Garcia96}%
  \BibitemOpen
  \bibfield  {author} {\bibinfo {author} {\bibfnamefont {V.~M.}\ \bibnamefont
  {P\'erez-Garc\'\i{}a}}, \bibinfo {author} {\bibfnamefont {H.}~\bibnamefont
  {Michinel}}, \bibinfo {author} {\bibfnamefont {J.~I.}\ \bibnamefont {Cirac}},
  \bibinfo {author} {\bibfnamefont {M.}~\bibnamefont {Lewenstein}}, \ and\
  \bibinfo {author} {\bibfnamefont {P.}~\bibnamefont {Zoller}},\ }\href@noop {}
  {\bibfield  {journal} {\bibinfo  {journal} {Phys. Rev. Lett.}\ }\textbf
  {\bibinfo {volume} {77}},\ \bibinfo {pages} {5320} (\bibinfo {year}
  {1996})}\BibitemShut {NoStop}%
\bibitem [{\citenamefont {Sinkin}\ \emph {et~al.}(2003)\citenamefont {Sinkin},
  \citenamefont {Holzl\"{o}hner}, \citenamefont {Zweck},\ and\ \citenamefont
  {Menyuk}}]{Sinkin03}%
  \BibitemOpen
  \bibfield  {author} {\bibinfo {author} {\bibfnamefont {O.~V.}\ \bibnamefont
  {Sinkin}}, \bibinfo {author} {\bibfnamefont {R.}~\bibnamefont
  {Holzl\"{o}hner}}, \bibinfo {author} {\bibfnamefont {J.}~\bibnamefont
  {Zweck}}, \ and\ \bibinfo {author} {\bibfnamefont {C.~R.}\ \bibnamefont
  {Menyuk}},\ }\href@noop {} {\bibfield  {journal} {\bibinfo  {journal} {J.
  Lightw. Techn.}\ }\textbf {\bibinfo {volume} {21}},\ \bibinfo {pages} {61}
  (\bibinfo {year} {2003})}\BibitemShut {NoStop}%
\bibitem [{\citenamefont {Stringari}(1996)}]{Stringari96}%
  \BibitemOpen
  \bibfield  {author} {\bibinfo {author} {\bibfnamefont {S.}~\bibnamefont
  {Stringari}},\ }\href@noop {} {\bibfield  {journal} {\bibinfo  {journal}
  {Phys. Rev. Lett.}\ }\textbf {\bibinfo {volume} {77}},\ \bibinfo {pages}
  {2360} (\bibinfo {year} {{1996}})}\BibitemShut {NoStop}%
\bibitem [{\citenamefont {Hall}\ \emph {et~al.}(2006)\citenamefont {Hall},
  \citenamefont {Whitlock}, \citenamefont {Scharnberg}, \citenamefont
  {Hannaford},\ and\ \citenamefont {Sidorov}}]{Hall06}%
  \BibitemOpen
  \bibfield  {author} {\bibinfo {author} {\bibfnamefont {B.~V.}\ \bibnamefont
  {Hall}}, \bibinfo {author} {\bibfnamefont {S.}~\bibnamefont {Whitlock}},
  \bibinfo {author} {\bibfnamefont {F.}~\bibnamefont {Scharnberg}}, \bibinfo
  {author} {\bibfnamefont {P.}~\bibnamefont {Hannaford}}, \ and\ \bibinfo
  {author} {\bibfnamefont {A.}~\bibnamefont {Sidorov}},\ }\href@noop {}
  {\bibfield  {journal} {\bibinfo  {journal} {J. Phys. B}\ }\textbf {\bibinfo
  {volume} {39}},\ \bibinfo {pages} {27} (\bibinfo {year} {2006})}\BibitemShut
  {NoStop}%
\bibitem [{\citenamefont {Anderson}\ \emph {et~al.}(2009)\citenamefont
  {Anderson}, \citenamefont {Ticknor}, \citenamefont {Sidorov},\ and\
  \citenamefont {Hall}}]{Anderson09}%
  \BibitemOpen
  \bibfield  {author} {\bibinfo {author} {\bibfnamefont {R.~P.}\ \bibnamefont
  {Anderson}}, \bibinfo {author} {\bibfnamefont {C.}~\bibnamefont {Ticknor}},
  \bibinfo {author} {\bibfnamefont {A.~I.}\ \bibnamefont {Sidorov}}, \ and\
  \bibinfo {author} {\bibfnamefont {B.~V.}\ \bibnamefont {Hall}},\ }\href@noop
  {} {\bibfield  {journal} {\bibinfo  {journal} {Phys. Rev. A}\ }\textbf
  {\bibinfo {volume} {80}},\ \bibinfo {pages} {023603} (\bibinfo {year}
  {2009})}\BibitemShut {NoStop}%
\bibitem [{\citenamefont {Ockeloen}\ \emph {et~al.}(2010)\citenamefont
  {Ockeloen}, \citenamefont {Tauschinsky}, \citenamefont {Spreeuw},\ and\
  \citenamefont {Whitlock}}]{Ockeloen10}%
  \BibitemOpen
  \bibfield  {author} {\bibinfo {author} {\bibfnamefont {C.~F.}\ \bibnamefont
  {Ockeloen}}, \bibinfo {author} {\bibfnamefont {A.~F.}\ \bibnamefont
  {Tauschinsky}}, \bibinfo {author} {\bibfnamefont {R.~J.~C.}\ \bibnamefont
  {Spreeuw}}, \ and\ \bibinfo {author} {\bibfnamefont {S.}~\bibnamefont
  {Whitlock}},\ }\href@noop {} {\bibfield  {journal} {\bibinfo  {journal}
  {Phys. Rev. A}\ }\textbf {\bibinfo {volume} {82}},\ \bibinfo {pages}
  {061606(R)} (\bibinfo {year} {2010})}\BibitemShut {NoStop}%
\bibitem [{\citenamefont {Widera}\ \emph {et~al.}(2006)\citenamefont {Widera},
  \citenamefont {Gerbier}, \citenamefont {F{\"{o}}lling}, \citenamefont
  {Gericke}, \citenamefont {Mandel},\ and\ \citenamefont {Bloch}}]{Widera06}%
  \BibitemOpen
  \bibfield  {author} {\bibinfo {author} {\bibfnamefont {A.}~\bibnamefont
  {Widera}}, \bibinfo {author} {\bibfnamefont {F.}~\bibnamefont {Gerbier}},
  \bibinfo {author} {\bibfnamefont {S.}~\bibnamefont {F{\"{o}}lling}}, \bibinfo
  {author} {\bibfnamefont {T.}~\bibnamefont {Gericke}}, \bibinfo {author}
  {\bibfnamefont {O.}~\bibnamefont {Mandel}}, \ and\ \bibinfo {author}
  {\bibfnamefont {I.}~\bibnamefont {Bloch}},\ }\href@noop {} {\bibfield
  {journal} {\bibinfo  {journal} {{New J. Phys.}}\ }\textbf {\bibinfo {volume}
  {{8}}},\ \bibinfo {pages} {152} (\bibinfo {year} {{2006}})}\BibitemShut
  {NoStop}%
\bibitem [{\citenamefont {Opanchuk}\ \emph {et~al.}(2012)\citenamefont
  {Opanchuk}, \citenamefont {Egorov}, \citenamefont {Hoffmann}, \citenamefont
  {Sidorov},\ and\ \citenamefont {Drummond}}]{Opanchuk11}%
  \BibitemOpen
  \bibfield  {author} {\bibinfo {author} {\bibfnamefont {B.}~\bibnamefont
  {Opanchuk}}, \bibinfo {author} {\bibfnamefont {M.}~\bibnamefont {Egorov}},
  \bibinfo {author} {\bibfnamefont {S.}~\bibnamefont {Hoffmann}}, \bibinfo
  {author} {\bibfnamefont {A.~I.}\ \bibnamefont {Sidorov}}, \ and\ \bibinfo
  {author} {\bibfnamefont {P.~D.}\ \bibnamefont {Drummond}},\ }\href@noop {}
  {\bibfield  {journal} {\bibinfo  {journal} {Eur. Phys. Lett.}\ }\textbf
  {\bibinfo {volume} {97}},\ \bibinfo {pages} {50003} (\bibinfo {year}
  {2012})}\BibitemShut {NoStop}%
\bibitem [{\citenamefont {Sinatra}\ \emph {et~al.}(2011)\citenamefont
  {Sinatra}, \citenamefont {Witkowska}, \citenamefont {Dornstetter},
  \citenamefont {Li},\ and\ \citenamefont {Castin}}]{Sinatra11}%
  \BibitemOpen
  \bibfield  {author} {\bibinfo {author} {\bibfnamefont {A.}~\bibnamefont
  {Sinatra}}, \bibinfo {author} {\bibfnamefont {E.}~\bibnamefont {Witkowska}},
  \bibinfo {author} {\bibfnamefont {J.-C.}\ \bibnamefont {Dornstetter}},
  \bibinfo {author} {\bibfnamefont {Y.}~\bibnamefont {Li}}, \ and\ \bibinfo
  {author} {\bibfnamefont {Y.}~\bibnamefont {Castin}},\ }\href@noop {}
  {\bibfield  {journal} {\bibinfo  {journal} {Phys. Rev. Lett.}\ }\textbf
  {\bibinfo {volume} {107}},\ \bibinfo {pages} {060404} (\bibinfo {year}
  {2011})}\BibitemShut {NoStop}%
\bibitem [{\citenamefont {Vogels}\ \emph {et~al.}(1997)\citenamefont {Vogels},
  \citenamefont {Tsai}, \citenamefont {Freeland}, \citenamefont {Kokkelmans},
  \citenamefont {Verhaar},\ and\ \citenamefont {Heinzen}}]{Vogels97}%
  \BibitemOpen
  \bibfield  {author} {\bibinfo {author} {\bibfnamefont {J.~M.}\ \bibnamefont
  {Vogels}}, \bibinfo {author} {\bibfnamefont {C.~C.}\ \bibnamefont {Tsai}},
  \bibinfo {author} {\bibfnamefont {R.~S.}\ \bibnamefont {Freeland}}, \bibinfo
  {author} {\bibfnamefont {S.~J. J. M.~F.}\ \bibnamefont {Kokkelmans}},
  \bibinfo {author} {\bibfnamefont {B.~J.}\ \bibnamefont {Verhaar}}, \ and\
  \bibinfo {author} {\bibfnamefont {D.~J.}\ \bibnamefont {Heinzen}},\
  }\href@noop {} {\bibfield  {journal} {\bibinfo  {journal} {Phys. Rev. A}\
  }\textbf {\bibinfo {volume} {56}},\ \bibinfo {pages} {R1067} (\bibinfo {year}
  {1997})}\BibitemShut {NoStop}%
\bibitem [{kok()}]{kokkelmans-private}%
  \BibitemOpen
  \href@noop {} {}\bibinfo {note} {S. J. J. M. F. Kokkelmans, private
  communication~(June 2010)}\BibitemShut {NoStop}%
\bibitem [{\citenamefont {Matthews}\ \emph {et~al.}(1998)\citenamefont
  {Matthews}, \citenamefont {Hall}, \citenamefont {Jin}, \citenamefont
  {Ensher}, \citenamefont {Wieman}, \citenamefont {Cornell}, \citenamefont
  {Dalfovo}, \citenamefont {Minniti},\ and\ \citenamefont
  {Stringari}}]{Matthews98}%
  \BibitemOpen
  \bibfield  {author} {\bibinfo {author} {\bibfnamefont {M.~R.}\ \bibnamefont
  {Matthews}}, \bibinfo {author} {\bibfnamefont {D.~S.}\ \bibnamefont {Hall}},
  \bibinfo {author} {\bibfnamefont {D.~S.}\ \bibnamefont {Jin}}, \bibinfo
  {author} {\bibfnamefont {J.~R.}\ \bibnamefont {Ensher}}, \bibinfo {author}
  {\bibfnamefont {C.~E.}\ \bibnamefont {Wieman}}, \bibinfo {author}
  {\bibfnamefont {E.~A.}\ \bibnamefont {Cornell}}, \bibinfo {author}
  {\bibfnamefont {F.}~\bibnamefont {Dalfovo}}, \bibinfo {author} {\bibfnamefont
  {C.}~\bibnamefont {Minniti}}, \ and\ \bibinfo {author} {\bibfnamefont
  {S.}~\bibnamefont {Stringari}},\ }\href@noop {} {\bibfield  {journal}
  {\bibinfo  {journal} {{Phys. Rev. Lett.}}\ }\textbf {\bibinfo {volume}
  {{81}}},\ \bibinfo {pages} {243} (\bibinfo {year} {{1998}})}\BibitemShut
  {NoStop}%
\bibitem [{\citenamefont {Li}\ \emph {et~al.}(2009)\citenamefont {Li},
  \citenamefont {Treutlein}, \citenamefont {Reichel},\ and\ \citenamefont
  {Sinatra}}]{Li09}%
  \BibitemOpen
  \bibfield  {author} {\bibinfo {author} {\bibfnamefont {Y.}~\bibnamefont
  {Li}}, \bibinfo {author} {\bibfnamefont {P.}~\bibnamefont {Treutlein}},
  \bibinfo {author} {\bibfnamefont {J.}~\bibnamefont {Reichel}}, \ and\
  \bibinfo {author} {\bibfnamefont {A.}~\bibnamefont {Sinatra}},\ }\href@noop
  {} {\bibfield  {journal} {\bibinfo  {journal} {{Eur.~Phys.~J.~B}}\ }\textbf
  {\bibinfo {volume} {{68}}},\ \bibinfo {pages} {365} (\bibinfo {year}
  {{2009}})}\BibitemShut {NoStop}%
\bibitem [{\citenamefont {Tacla}\ \emph {et~al.}(2010)\citenamefont {Tacla},
  \citenamefont {Boixo}, \citenamefont {Datta}, \citenamefont {Shaji},\ and\
  \citenamefont {Caves}}]{Tacla10}%
  \BibitemOpen
  \bibfield  {author} {\bibinfo {author} {\bibfnamefont {A.~B.}\ \bibnamefont
  {Tacla}}, \bibinfo {author} {\bibfnamefont {S.}~\bibnamefont {Boixo}},
  \bibinfo {author} {\bibfnamefont {A.}~\bibnamefont {Datta}}, \bibinfo
  {author} {\bibfnamefont {A.}~\bibnamefont {Shaji}}, \ and\ \bibinfo {author}
  {\bibfnamefont {C.~M.}\ \bibnamefont {Caves}},\ }\href@noop {} {\bibfield
  {journal} {\bibinfo  {journal} {Phys. Rev. A}\ }\textbf {\bibinfo {volume}
  {82}},\ \bibinfo {pages} {053636} (\bibinfo {year} {2010})}\BibitemShut
  {NoStop}%
\bibitem [{\citenamefont {Tojo}\ \emph {et~al.}(2010)\citenamefont {Tojo},
  \citenamefont {Taguchi}, \citenamefont {Masuyama}, \citenamefont {Hayashi},
  \citenamefont {Saito},\ and\ \citenamefont {Hirano}}]{Tojo10}%
  \BibitemOpen
  \bibfield  {author} {\bibinfo {author} {\bibfnamefont {S.}~\bibnamefont
  {Tojo}}, \bibinfo {author} {\bibfnamefont {Y.}~\bibnamefont {Taguchi}},
  \bibinfo {author} {\bibfnamefont {Y.}~\bibnamefont {Masuyama}}, \bibinfo
  {author} {\bibfnamefont {T.}~\bibnamefont {Hayashi}}, \bibinfo {author}
  {\bibfnamefont {H.}~\bibnamefont {Saito}}, \ and\ \bibinfo {author}
  {\bibfnamefont {T.}~\bibnamefont {Hirano}},\ }\href@noop {} {\bibfield
  {journal} {\bibinfo  {journal} {Phys. Rev. A}\ }\textbf {\bibinfo {volume}
  {82}},\ \bibinfo {pages} {033609} (\bibinfo {year} {2010})}\BibitemShut
  {NoStop}%
\bibitem [{\citenamefont {Nicklas}\ \emph {et~al.}(2011)\citenamefont
  {Nicklas}, \citenamefont {Strobel}, \citenamefont {Zibold}, \citenamefont
  {Gross}, \citenamefont {Malomed}, \citenamefont {Kevrekidis},\ and\
  \citenamefont {Oberthaler}}]{Niclas11}%
  \BibitemOpen
  \bibfield  {author} {\bibinfo {author} {\bibfnamefont {E.}~\bibnamefont
  {Nicklas}}, \bibinfo {author} {\bibfnamefont {H.}~\bibnamefont {Strobel}},
  \bibinfo {author} {\bibfnamefont {T.}~\bibnamefont {Zibold}}, \bibinfo
  {author} {\bibfnamefont {C.}~\bibnamefont {Gross}}, \bibinfo {author}
  {\bibfnamefont {B.~A.}\ \bibnamefont {Malomed}}, \bibinfo {author}
  {\bibfnamefont {P.~G.}\ \bibnamefont {Kevrekidis}}, \ and\ \bibinfo {author}
  {\bibfnamefont {M.~K.}\ \bibnamefont {Oberthaler}},\ }\href@noop {}
  {\bibfield  {journal} {\bibinfo  {journal} {Phys. Rev. Lett.}\ }\textbf
  {\bibinfo {volume} {107}},\ \bibinfo {pages} {193001} (\bibinfo {year}
  {2011})}\BibitemShut {NoStop}%
\bibitem [{\citenamefont {Li}\ \emph {et~al.}(2008)\citenamefont {Li},
  \citenamefont {Castin},\ and\ \citenamefont {Sinatra}}]{Li08}%
  \BibitemOpen
  \bibfield  {author} {\bibinfo {author} {\bibfnamefont {Y.}~\bibnamefont
  {Li}}, \bibinfo {author} {\bibfnamefont {Y.}~\bibnamefont {Castin}}, \ and\
  \bibinfo {author} {\bibfnamefont {A.}~\bibnamefont {Sinatra}},\ }\href@noop
  {} {\bibfield  {journal} {\bibinfo  {journal} {Phys. Rev. Lett.}\ }\textbf
  {\bibinfo {volume} {100}},\ \bibinfo {pages} {210401} (\bibinfo {year}
  {2008})}\BibitemShut {NoStop}%
\bibitem [{\citenamefont {Kaufman}\ \emph {et~al.}(2009)\citenamefont
  {Kaufman}, \citenamefont {Anderson}, \citenamefont {Hanna}, \citenamefont
  {Tiesinga}, \citenamefont {Julienne},\ and\ \citenamefont
  {Hall}}]{Kaufman09}%
  \BibitemOpen
  \bibfield  {author} {\bibinfo {author} {\bibfnamefont {A.~M.}\ \bibnamefont
  {Kaufman}}, \bibinfo {author} {\bibfnamefont {R.~P.}\ \bibnamefont
  {Anderson}}, \bibinfo {author} {\bibfnamefont {T.~M.}\ \bibnamefont {Hanna}},
  \bibinfo {author} {\bibfnamefont {E.}~\bibnamefont {Tiesinga}}, \bibinfo
  {author} {\bibfnamefont {P.~S.}\ \bibnamefont {Julienne}}, \ and\ \bibinfo
  {author} {\bibfnamefont {D.~S.}\ \bibnamefont {Hall}},\ }\href@noop {}
  {\bibfield  {journal} {\bibinfo  {journal} {{Phys. Rev. A}}\ }\textbf
  {\bibinfo {volume} {{80}}},\ \bibinfo {pages} {050701} (\bibinfo {year}
  {{2009}})}\BibitemShut {NoStop}%
\bibitem [{\citenamefont {Tscherbul}\ \emph {et~al.}(2010)\citenamefont
  {Tscherbul}, \citenamefont {Calarco}, \citenamefont {Lesanovsky},
  \citenamefont {Krems}, \citenamefont {Dalgarno},\ and\ \citenamefont
  {Schmiedmayer}}]{Tscherbul10}%
  \BibitemOpen
  \bibfield  {author} {\bibinfo {author} {\bibfnamefont {T.~V.}\ \bibnamefont
  {Tscherbul}}, \bibinfo {author} {\bibfnamefont {T.}~\bibnamefont {Calarco}},
  \bibinfo {author} {\bibfnamefont {I.}~\bibnamefont {Lesanovsky}}, \bibinfo
  {author} {\bibfnamefont {R.~V.}\ \bibnamefont {Krems}}, \bibinfo {author}
  {\bibfnamefont {A.}~\bibnamefont {Dalgarno}}, \ and\ \bibinfo {author}
  {\bibfnamefont {J.}~\bibnamefont {Schmiedmayer}},\ }\href@noop {} {\bibfield
  {journal} {\bibinfo  {journal} {Phys. Rev. A}\ }\textbf {\bibinfo {volume}
  {81}},\ \bibinfo {pages} {050701} (\bibinfo {year} {2010})}\BibitemShut
  {NoStop}%
\bibitem [{\citenamefont {Papoular}\ \emph {et~al.}(2010)\citenamefont
  {Papoular}, \citenamefont {Shlyapnikov},\ and\ \citenamefont
  {Dalibard}}]{Papoular10}%
  \BibitemOpen
  \bibfield  {author} {\bibinfo {author} {\bibfnamefont {D.~J.}\ \bibnamefont
  {Papoular}}, \bibinfo {author} {\bibfnamefont {G.~V.}\ \bibnamefont
  {Shlyapnikov}}, \ and\ \bibinfo {author} {\bibfnamefont {J.}~\bibnamefont
  {Dalibard}},\ }\href@noop {} {\bibfield  {journal} {\bibinfo  {journal}
  {Phys. Rev. A}\ }\textbf {\bibinfo {volume} {{81}}},\ \bibinfo {pages}
  {041603} (\bibinfo {year} {{2010}})}\BibitemShut {NoStop}%
\bibitem [{\citenamefont {Chin}\ and\ \citenamefont {Flambaum}(2006)}]{Chin06}%
  \BibitemOpen
  \bibfield  {author} {\bibinfo {author} {\bibfnamefont {C.}~\bibnamefont
  {Chin}}\ and\ \bibinfo {author} {\bibfnamefont {V.~V.}\ \bibnamefont
  {Flambaum}},\ }\href@noop {} {\bibfield  {journal} {\bibinfo  {journal}
  {Phys. Rev. Lett.}\ }\textbf {\bibinfo {volume} {96}},\ \bibinfo {pages}
  {230801} (\bibinfo {year} {2006})}\BibitemShut {NoStop}%
\end{thebibliography}%

\end{document}